\documentclass[aps,prx,twocolumn,final,letterpaper]{revtex4}

\usepackage{appendix}
\usepackage{graphicx}   
\usepackage{import}                         
\usepackage{epstopdf}
\usepackage{bm}
\usepackage{amssymb}
\usepackage{quotes}
\usepackage{transparent}
\usepackage{dcolumn}
\usepackage{multirow}
\usepackage{cancel} 
\usepackage{mdframed}
\usepackage{color}
\usepackage{bm}
\usepackage{dsfont}
\usepackage{slashed}
\usepackage{enumitem}
\usepackage{amsmath} 
\usepackage{comment}
\begin{document}
\rmfamily

\title{Strong broadband intensity noise squeezing from near-infrared to terahertz frequencies in semiconductor lasers with nonlinear dissipation}

\author{Sahil Pontula$^{1,2,4,\dagger}$, Jamison Sloan$^{1,2,\dagger}$, Nicholas Rivera$^{1,3}$, and Marin Solja\v{c}i\'{c}$^{1,4}$}

\affiliation{$^{1}$Department of Physics, Massachusetts Institute of Technology, Cambridge, MA 02139, USA. \\
$^{2}$Department of Electrical Engineering and Computer Science, Massachusetts Institute of Technology, Cambridge, MA 02139, USA.  \\
$^{3}$Department of Physics, Harvard University, Cambridge, MA 02138, USA. \\
$^{4}$Research Laboratory of Electronics, Massachusetts Institute of Technology, Cambridge, MA 02139, USA. \\
$\dagger$ Denotes equal contribution. }

\begin{abstract} 


The generation and application of squeezed light have long been central goals of quantum optics, enabling sensing below the standard quantum limit, optical quantum computing platforms, and more. Intensity noise squeezing of bright (coherent) states, in contrast to squeezed vacuum, is relatively underdeveloped. Bright squeezing has been generated directly through nonlinear optical processes or ``quietly pumped'' semiconductor lasers. However, these methods suffer from weak squeezing limits, narrow operating wavelength ranges, and have not been explored at large bandwidths. Here, we show how semiconductor lasers with sharp intensity-dependent dissipation can support highly broadband intensity noise squeezing from infrared (IR) to terahertz (THz) wavelengths, the latter of which has remained unexplored in quantum noise studies. Our protocol realizes strongly ($>10$ dB) intensity noise-squeezed intracavity quantum states, which could create a new regime for cavity quantum electrodynamics experiments, as well as strong output squeezing surpassing gigahertz bandwidths. Furthermore, we show how the same systems also create self-pulsing and bistable mean field behavior, enabling control of light in both the temporal and noise domains. The existence of these multiple functionalities in both the classical and quantum mechanical domains in a single semiconductor laser platform, from IR to THz wavelengths, could enable advances in on-chip quantum optical communication, computing, and sensing across the electromagnetic spectrum.


\end{abstract}

\maketitle

\section{Introduction}

The generation of states of light with noise ``squeezed'' below the standard quantum limit for a coherent state is a decades-old pursuit of quantum optics. In these squeezed states, the variance in one observable (such as amplitude or phase) is reduced at the expense of another, permitting levels of quantum fluctuations which lie below the standard quantum limit. Such squeezed states of light have been harnessed for optical quantum computing as well as precision sensing and metrology \cite{lawrie2019quantum, madsen2022quantum}. The most common methods to generate squeezed light employ laser-pumped nonlinear crystals. For example, sub-threshold optical parametric amplifiers have been used to produce up to 3 dB of intracavity squeezing \cite{collett1984squeezing,dassonneville2021dissipative, qin2022beating} and 15 dB of propagating squeezed vacuum \cite{wu1986generation,vahlbruch2008observation,vahlbruch2016detection}. 

By contrast, schemes to generate squeezing in bright states of light are less mature, despite their promise as sources for sensitive spectroscopy applications and pumps for low-noise optical amplifiers \cite{polzik1992spectroscopy,collett1988quantum}. Generation of bright squeezing has been limited to methods developed over two decades ago such as second harmonic generation, Kerr nonlinearity in fiber-optic interferometers, and ``quietly pumped'' semiconductor lasers \cite{yamamoto1992photon, paschotta1994bright, schmitt1998photon}. However, these mechanisms come with inherent tradeoffs that limit the space of possible applications. First, the magnitude of squeezing achieved has not approached that achievable with squeezed vacuum (in either intracavity or output intensity noise), limiting applications where intense squeezed light is preferred over squeezed vacuum. Secondly, large (GHz) bandwidths have not been demonstrated with these bright squeezing methods, limiting their application in quantum communication protocols. In mesoscopic systems with strong nonlinearities, high levels of broadband intracavity squeezing could produce approximate large Fock states, with exciting potential applications in qubit nondemolition readout in cavity QED, optomechanical cooling, quantum metrology, and enhanced light-matter interactions \cite{turchette1998squeezed, wallquist2010single, peano2015intracavity, gan2019intracavity, qin2022beating}. Finally, existing methods to produce intense squeezed states have been generally limited to narrow wavelength ranges in the infrared (e.g., due to nonlinear phase matching and conversion efficiency constraints). As a result, there are large wavelength ranges (MIR-THz) in which intensity squeezing has never been demonstrated, despite tantalizing applications in quantum-enhanced chemical fingerprinting, wireless communication, and solid-state qubit manipulation \cite{todorov2024thz}. 


These wavelengths spanning from the IR to the THz have been particularly well-served by semiconductor lasers, owing to their wide gain bandwidths, convenient form factors, and ease of electrical pumping. Several methods have been explored to produce intensity squeezing directly from semiconductor lasers, including so-called  ``quiet pumping'' (pump noise suppression) and optical feedback/dispersive loss to exploit amplitude-phase correlations \cite{newkirk1991amplitude, kitching1995room, jeremie1999room}. However, these methods have achieved only a few dB of squeezing. Moreover, such squeezing has been achieved only at low noise frequencies, leaving the large excess noise from so-called ``relaxation oscillations'' at higher frequencies unmitigated. Thus, the majority of modern semiconductor lasers do not surpass --- or even reach --- the shot noise limit at large bandwidths. This, together with the limitations of other nonlinear optical techniques described above, highlights a broad open challenge in producing sources of highly squeezed intense light which are versatile in wavelength and bandwidth.

Here, we show how semiconductor lasers equipped with Kerr nonlinearity and frequency-dependent outcoupling can enable sharply nonlinear dissipation and act as a source of intense squeezed light from IR to THz wavelengths, reducing intracavity fluctuations to more than 10 dB below the shot noise limit. Output fluctuations are significantly suppressed relative to conventional semiconductor lasers and, when combined with quiet pumping schemes, can be squeezed over 10 dB below the shot noise limit at GHz bandwidths. Our approach exploits intensity-dependent dissipation, in conjunction with a semiconductor gain medium, to create a laser architecture which natively produces light with intensity fluctuations far below the shot noise limit. We show that semiconductor laser architectures are aptly suited for this purpose due to their compact form factor, strong intrinsic optical nonlinearities, and ease of on-chip integration with the low loss resonators and photonic crystals required to generate frequency-dependent dissipation. In addition, we explain how these same architectures can exhibit classical nonlinear phenomena such as self-pulsing and bistability. Together, these functionalities could pave the way towards combined temporal and quantum noise control over light across the electromagnetic spectrum. This could unlock elusive quantum states such as THz pulsed squeezed states, with novel applications in communications and sensing. 

We structure the rest of this paper as follows. Section \ref{sec:theory} shows how the concept of intensity-dependent dissipation can be implemented in a semiconductor laser architecture. Section \ref{sec:results} first presents the mean-field dynamics of a semiconductor laser with nonlinear dispersive loss, demonstrating self-pulsing and bistability, and then strong, broadband intracavity and output intensity noise squeezing at IR and THz wavelengths. Finally, Sections \ref{sec:discussion}, \ref{sec:outlook} summarize our results and present outlooks and future extensions of this work.

\section{Theory} 
\label{sec:theory}

\begin{figure}[t]
    \centering
    \includegraphics[scale=0.7]{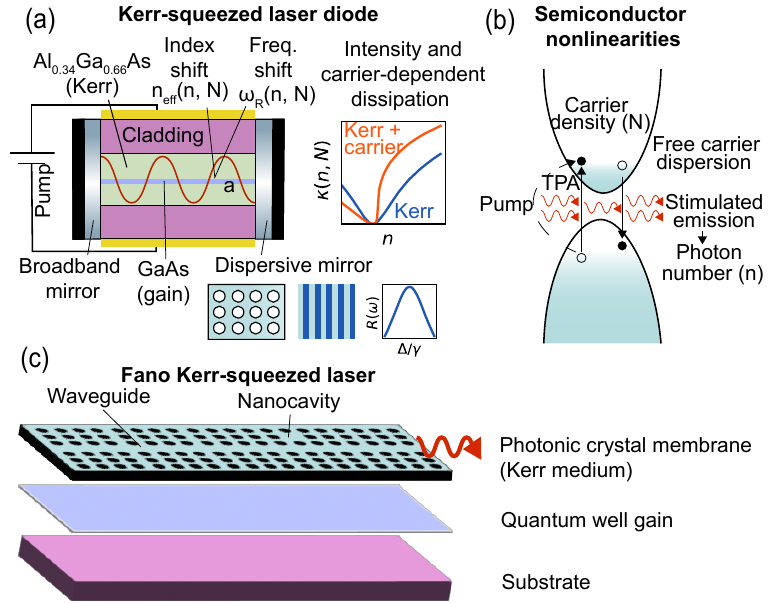}
    \caption{\textbf{Semiconductor lasers with nonlinear dispersive loss.} (a) Basic semiconductor laser diode heterostructure design with nonlinear dispersive loss. Dispersive outcoupling is generated via the sharp frequency-dependent transmission of a photonic crystal element. Coupling of Kerr nonlinearity from the Kerr material and carrier nonlinearity from the gain material with a dispersive mirror of reflectivity $R(\omega)$ creates sharp nonlinear loss $\kappa(n,N)$. Here, $\Delta$ denotes detuning from the dispersive (Lorentzian) resonance and $\gamma$ denotes the width of the dispersive resonance (related to its FWHM). (b) Semiconductor optical nonlinearities, including carrier-dependent free carrier dispersion (FCD) and two photon absorption (TPA). In addition to the photon number-dependent Kerr effect, these nonlinearities shift the real part of the active region's refractive index, in turn shifting the resonance frequency in the laser cavity. Weak nonlinear loss from shifting the imaginary part of the refractive index via the Kramers-Kronig relations is also generated, but in most cases is negligible compared to the nonlinear dispersive loss. (c) Sample implementation of nonlinear loss in a photonic crystal (PhC) ``Fano'' laser. The PhC platform allows much stronger per-photon nonlinearities due to very small mode volumes. Dispersive loss is provided by waveguide-nanocavity Fano interference in a photonic crystal slab \cite{yu2017demonstration}.}
    \label{fig:fock_schema}
\end{figure}

\subsection{Nonlinear dispersive loss}

We first describe how, under the right conditions, the combination of Kerr nonlinearity and frequency-dependent loss lead to a laser cavity with an effective \emph{intensity-dependent loss} that controls the quantum state of light produced by the laser. Consider the cavity architecture shown for a semiconductor laser in Fig. 1a. We focus on a single cavity mode, with annihilation operator $a$. As is well known, a cavity containing a Kerr nonlinearity develops an intensity-dependent resonance frequency due to the intensity-dependent index of the Kerr material \cite{drummond1980quantum}. In the case of semiconductor lasers, free carrier nonlinearities (Fig. \ref{fig:fock_schema}b) also shift the cavity resonance. Then, the cavity resonance frequency depends linearly on the photon number and inverted carrier density $n$ and $N$ as
\begin{align}
    \omega_R(n,N)=\omega_0\cdot (1+\beta n+\sigma N),
\end{align}
as derived in the S.I.. This form for the cavity resonance shift due to semiconductor nonlinearities has been analyzed previously using coupled mode theory and supported experimentally \cite{heuck2013heterodyne, yu2013switching, lunnemann2012nonlinear, said1992determination}. Here, $\omega_0$ is the bare resonance frequency of the cavity mode $a$, $\beta$ is a dimensionless per-photon nonlinearity that can be directly calculated from the Kerr nonlinear coefficient $n_2$ or nonlinear susceptibility $\chi^{(3)}$, and the carrier nonlinearity $\sigma$ is material-dependent and is directly related to the linewidth enhancement factor (see S.I. for details). 

Additionally, in the laser cavity of Fig. \ref{fig:fock_schema}a, one of the end facets is a broadband reflector, while the other is a sharply dispersive element, such as a Fano resonance structure or a Bragg reflector, which equips the cavity with sharply frequency-dependent dissipation through its reflection coefficient $R(\omega)$. When combined, the intensity-dependent resonance frequency and frequency-dependent dissipation give the cavity mode an effective \emph{intensity-dependent dissipation}, which can promote the formation of quantum states \cite{A:V1,A:V2}. The one critical assumption for this description is that the temporal response of the dispersive mirror is fast compared to the round trip time of the cavity. This corresponds to an adiabatic limit where the dispersive resonance, which sets the cavity transmission $T(\omega)$, is able to near-instantaneously follow shifts in the cavity frequency caused by the nonlinearities. When these assumptions are fulfilled, the cavity field is subject to an effective intensity-dependent damping rate
\begin{align}
\begin{split}
            \kappa(n,N)\equiv \kappa(\omega_R(n,N)) &= -\mathrm{FSR}\cdot \log{R(\omega_R(n,N))}\\
            &\approx \mathrm{FSR}\cdot T(\omega_R(n,N)),
    \label{eq:loss}
\end{split}
\end{align}
where the approximation holds when $R(\omega_R)\approx 1$. Sharply frequency-dependent reflectivity profiles enable the dissipation rate $\kappa(n,N)$ to take on forms which are highly nonperturbative in $n$, making this type of nonlinear dissipation fundamentally different than the types of nonlinear dissipation realized by multi-photon absorption. One example of such a reflectivity profile has been realized in self-pulsing Fano lasers \cite{yu2017demonstration} with low mode volumes which, when augmented with a Kerr nonlinear material (Fig. \ref{fig:fock_schema}c), could create strongly nonlinear dissipation. As we will show, systems exhibiting this kind of loss can provide new behaviors not just in their steady states, but also through new quantum noise behaviors.

Note that in Fig. \ref{fig:fock_schema}a we consider a semiconductor laser with separate gain and Kerr nonlinear elements. We choose to use a different material for the Kerr nonlinearity in order to avoid possible dispersive resonant effects of optical nonlinearity near transition energies in the gain material. The Kerr material is chosen to be a GaAs-based semiconductor due to its strong optical nonlinearity from bound carriers. Semiconductor lasers with nonlinear dispersive loss based on ``active nonlinearity'' (in which the gain and Kerr materials are the same) may be possible, but the timescale of resonant effects may call into question the adiabatic assumption of the cavity resonance frequency's instantaneous response to changes in photon number, thus placing such systems outside the scope of the models we consider here.

\subsection{Laser dynamics}
\label{sec:archs}

Semiconductors typically fall into the category of so-called ``class B'' lasers, in which the polarization dynamics decay quickly relative to the timescales associated with carrier recombination and cavity decay. In this case, the polarization dynamics are adiabatically eliminated, resulting in Heisenberg-Langevin equations for photon number and carrier number operators, as derived in the S.I. \cite{agrawal2012fiber, chow2012semiconductor}: 
\begin{subequations}
\begin{align}
\dot n &=\left(G(n,N) - \kappa(n,N)\right)n+F_n \\
\dot N &=I - \left(nG(n,N) + \gamma_\parallel N\right)+F_N.
\end{align}
\label{eq:rates}
\end{subequations}
To be maximally general here, we allow the gain $G$ and loss $\kappa$ to depend on both the carrier density $N$ and photon number $n$ (the latter could account for gain saturation). In writing this form of the gain and loss, we have assumed that the gain and loss respond effectively instantaneously to changes in the photon and carrier number. Pumping is performed by carrier injection using current $I$ (in units of carrier density per unit time), and $\gamma_\parallel$ denotes the nonradiative decay rate of carriers. The case of optically pumped excitation of free carriers is described in the S.I.. Finally, the decay rates and pump noise are associated with zero-mean Langevin force terms $F_{n,N,}$ with nonzero correlators provided in the S.I..

In all examples presented in the main text, we consider linear gain which neglects saturation effects, so that $G(n,N)=G(N)=G_N(N-N_{\mathrm{trans}})$ with $N_{\mathrm{trans}}$ the transparency carrier density. We found no phenomenological differences using logarithmic quantum well gain or including the effects of gain saturation \cite{agrawal1990effect}. 

\subsection{Noise properties}

\begin{figure*}[t]
    \centering
    \includegraphics[scale=0.7]{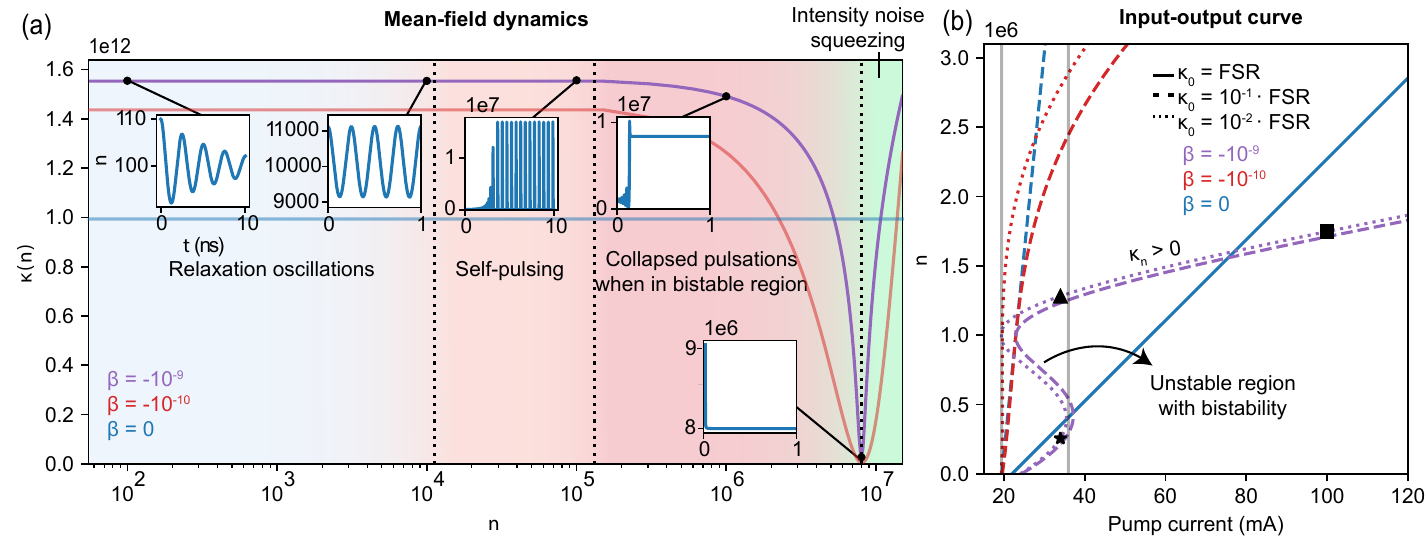}
    \caption{\textbf{Mean-field dynamics and steady state behavior.} (a) Dynamical and steady state solutions in semiconductor lasers with nonlinear dispersive loss. In the region $\kappa_n=\partial \kappa/\partial n<0$ (photon numbers left of the Fano resonance), a variety of different behaviors are possible. At large detunings (small $n$, blue region), the loss does not depend strongly on photon number, and the relaxation oscillations typical of conventional semiconductor lasers are observed. At a certain detuning ($n$), the relaxation oscillations become critically damped and, at smaller detunings, they become undamped, leading to self-sustained picosecond pulses (orange region). When the pump enters the bistable region (red region, only present for $\beta=-10^{-9}$ (purple curve)), the pulses become transient and the laser ultimately collapses to a continuous wave (CW) steady state. Lastly, to the right of the loss minimum (green), relaxation oscillations are heavily damped since $\kappa_n=\partial\kappa/\partial n>0$, leaving a CW steady state. Plots were produced by considering a transient increase in intracavity intensity by 10\% at $t=0$ relative to steady state. (b) Steady state intracavity photon number $n$ as a function of pump current (S-curve) for three different linear background losses $\kappa_0$ and nonlinear strengths $\beta$. The indicated unstable region is bypassed by the bistable point and is not generally accessible during lasing. The gray vertical lines denote the boundaries of the bistable region for the purple dotted curve. In these simulations, we use parameters based on experimentally determined values for buried heterostructure lasers with GaAs gain and AlGaAs cladding (Fig. \ref{fig:fock_schema}a): active region dimensions $0.1$ $\mu$m $\times$ $5$ $\mu$m $\times$ $1$ mm, confinement factor $\Gamma=0.3$, bare cavity resonance frequency $\omega_0=2.16\times 10^{15}$ s\textsuperscript{-1} (873 nm, GaAs bandgap), free spectral range $\mathrm{FSR}=43$ GHz, transparency density $N_{\mathrm{trans}}=2\times 10^{24}$ m\textsuperscript{-3}, nonradiative decay rate $\gamma_\parallel=3\times 10^8$ s\textsuperscript{-1}, and linear gain coefficient $G_N=1/V\cdot dG/dN=3694$ s\textsuperscript{-1} \cite{chow2012semiconductor}. The S.I. provides an estimate of typical Kerr nonlinear strengths in this structure. The Fano resonance is centered at photon number $n_c=8\times 10^6$ (a) and $n_c=10^6$ (b). Its resonance decay (FWHM) is $\gamma=2\times 10^{12}$ s\textsuperscript{-1}. In (a), $\kappa_0=10^{-2}\cdot \mathrm{FSR}$ for $\beta\ne 0$.}
    \label{fig:f2}
\end{figure*}

The noise properties of semiconductor lasers can be computed by considering operator valued fluctuations of the Heisenberg-Langevin equations from their mean field solutions. In the steady state, this results in a pair of coupled linear equations for the operator values fluctuations $\delta n$ and $\delta N$, which are given as:

\begin{align}
\begin{bmatrix}
\delta \dot n \\
\delta\dot N
\end{bmatrix}
= \begin{bmatrix}
-n\kappa_n & n\left(G_N-\kappa_N\right)\\
-G_0 & -(nG_N+\gamma_\parallel)
\end{bmatrix}
\begin{bmatrix}
\delta n\\
\delta N
\end{bmatrix} 
+ 
\begin{bmatrix}
F_n-n\kappa_\omega F_\phi\\
F_N
\end{bmatrix}.    
\label{eq:fluc}
\end{align}

\noindent Here, $\kappa_{n}\equiv \partial\kappa/\partial n=-\beta\omega_0\kappa_\omega = -\beta\omega_0 (\partial\kappa/\partial\omega)$ represents the sharpness of the dispersive loss, $\kappa_N\equiv \partial\kappa/\partial N=\alpha_LG_N\kappa_{\omega}/2=-\sigma\omega_0\kappa_\omega$, where $\alpha_L$ is the linewidth enhancement factor (directly related to $\sigma$, the free carrier dependence of the refractive index), and $F_\phi$ is a Langevin force associated with the phase equation of motion. Note that $\alpha_L$, which emerges due to amplitude-phase coupling in semiconductor lasers, affects noise behavior, but not steady state operation. Important physical parameters to characterize intensity noise are the relaxation oscillation frequency and damping rate of relaxation oscillations. As derived in the S.I., these can be calculated from the complex poles of the fluctuation dynamics: 
\begin{align}
\begin{split}
    \Omega_R^2 &\approx (nG_N+\gamma_\parallel)(n\kappa_n)+n(G_N-\kappa_N)\kappa \\
\Gamma_1 &\approx n(G_N+\kappa_n)+\gamma_\parallel.
\end{split}
\label{eq:ro}
\end{align}
These measures provide an important way to understand the effect of nonlinear dispersive loss on quantum noise. They will also dictate mean field dynamics that result from fluctuations from steady state operation.

Going forward, we will assume in the main text that the shift in refractive index due to Kerr nonlinearity is much stronger than that due to carrier nonlinearity, $|\beta n|\gg |\sigma(N-N_\mathrm{trans})|$, so that the dependence of loss on carrier number $\kappa_N$ can be neglected. With strong Kerr nonlinearity, this is generally true for linewidth enhancement factors $\alpha_L\lesssim 5$. Many semiconductor laser systems fall in this regime, but quantum well/quantum dot designs and gain-symmetric quantum cascade lasers generally minimize $\alpha_L$ \cite{ukhanov2004comparison,green2008linewidth}. We consider the behavior when $|\beta n|\sim |\sigma (N-N_\mathrm{trans})|$ as well as the effect of two photon absorption in the S.I..

\section{Results}
\label{sec:results}


\subsection{Mean-field dynamics}
\label{sec:mf}

\begin{figure*}[t]
    \centering
    \includegraphics[scale=0.7]{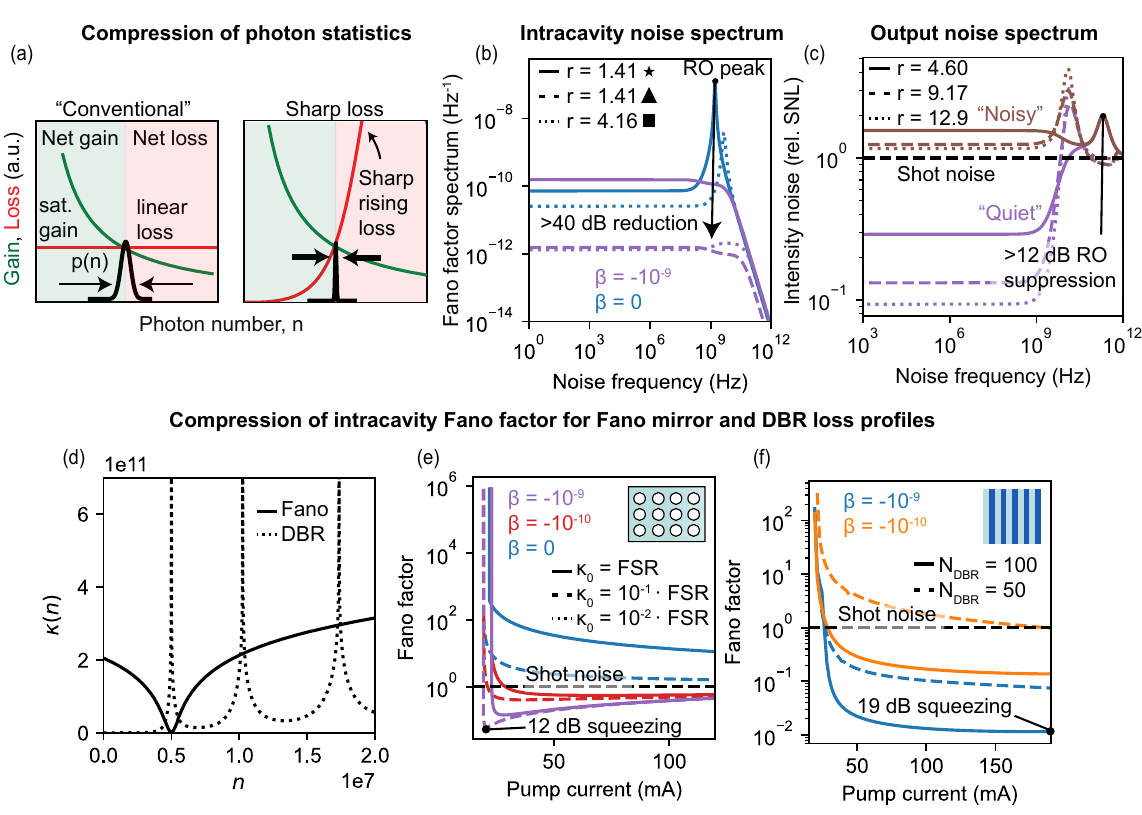}
    \caption{\textbf{Intensity noise squeezing.} (a) Comparison of steady state photon probability distribution $p(n)$ under conventional and sharp loss. The steady state photon number is determined by the location of intersection between saturable gain and loss. The variance of the probability distribution is determined by the effective ``steepness'' of intersection of the gain and loss curves. While the conventional laser architecture with linear loss results in a near-coherent state far above threshold, the sharp loss architecture results in states with variance below the mean, which correspond to non-classical light. In the most extreme limit, this mechanism can enable the generation of near-Fock states inside the laser cavity. (b) Intracavity Fano factor spectrum ($\Delta n^2(\omega)/n$) as a function of noise frequency for the three different steady states ($\bigstar, \blacktriangle, \blacksquare$) indicated in the input-output curve of Fig. \ref{fig:f2}b ($r\equiv I/I_\mathrm{thres}$ is the pumping ratio). Here, $\kappa_0=\mathrm{FSR}$ for the linear loss (blue) and $\kappa_0=10^{-2}\cdot\mathrm{FSR}$ for the nonlinear loss (purple). Nonlinear loss creates a strong suppression of the relaxation oscillation (RO) peak. (c) Output squeezing over a $>1$ GHz bandwidth with (``noisy'') and without (``quiet'') pump noise suppression (plotted for three different pump powers with nonlinear strength $\beta=-10^{-9}$). (d), (e), (f) Comparison of loss profiles and integrated Fano factor as a function of pump current for a nonlinear laser with a Fano mirror or DBR. Fano factors are plotted for the low noise branch when bistability is present. In (d), (e), (f), $n_c=5\times 10^6$ marks the center of the Fano resonance, while for DBR loss profiles, the average index is $\tilde n = 3.0$, the index contrast is $\Delta n\lesssim 1.0$, and the first transition from stop to pass band is tuned to occur around $n_c=5\times 10^6$.  All other simulation parameters are the same as those in Fig. \ref{fig:f2}.}
    \label{fig:f3}
\end{figure*}

We begin by considering the mean-field steady-state and dynamical solutions that emerge for a Kerr nonlinear semiconductor laser with a symmetric Fano resonance. As shown in Fig. \ref{fig:f2}a, the mean-field behavior can differ drastically depending on the sharpness of the loss $\kappa_n=\partial\kappa/\partial n$.

The mean-field dynamics of the equations of motion allow diverse modes of operation, as shown in Fig. \ref{fig:f2}a. The key driving force for these behaviors is the variation in the damping rate for relaxation oscillations (Eq. \ref{eq:ro}), which describes relaxation back to the mean field steady state. We plot the temporal evolution of the intracavity photon number following a transient 10\% increase in the photon number at $t=0$ relative to the initial steady state. For $\kappa_n\approx 0$ (low $n$ and far detuned from Fano resonance, blue region), relaxation oscillations are observed. For $\kappa_n\ll 0$, the relaxation oscillations become critically damped and eventually undamped (orange region), resulting in oscillations that transition into self-generated and self-sustained pulses. The pulses are quenched when the initial photon number enters the bistable region's lowest branch (demarcated by gray lines in Fig. \ref{fig:f2}b), ultimately collapsing to the topmost branch (and bypassing the intermediate unstable branch). For $\kappa_n>0$, relaxation oscillations are strongly damped ($\Gamma_1$ grows with $\kappa_n$ in Eq. \ref{eq:ro}). We note that in many semiconductor lasers that are not operated very far above threshold, intensity noise is often far from the shot noise limit due to the relaxation oscillation peak. The nonlinear loss in the region $\kappa_n>0$ suppresses this peak by over four orders of magnitude, as we will show. Physically, the nonlinear loss magnifies the strength of attraction of the laser steady state (a fixed point of the rate equations) in proportion to the slope $\kappa_n$. This has the effect of strongly resisting deviations from the steady state photon number, leading to the strong intensity squeezing described in the next section.

Self-pulsing has been reported previously using photonic crystal-based ``Fano'' lasers with saturable free carrier absorption from a nanocavity \cite{yu2017demonstration}. Here, we see that a similar phenomenon occurs due to a different physical mechanism: the combination of Kerr nonlinearity and dispersive loss. Suppose that the laser is pumped to a CW steady-state lying in the self-pulsing region of Fig. \ref{fig:f2}a. A transient increase in intracavity intensity (e.g., due to spontaneous emission into the lasing mode) now decreases the photonic loss, providing a positive feedback mechanism that builds up the intracavity intensity further. This should continue up to the point where the stimulated emission rate is high enough to drop the carrier density below threshold. The pump then builds up the carrier density again, and the pulsing continues. Further details about the self-pulsing behavior, including an analysis of the pulse profile, are provided in the S.I..

 We now examine the steady-state input-output curve (S-curve), as shown in Fig. \ref{fig:f2}b. Linear loss presents an $n$-independent loss profile, and leads to the well known linear dependence of steady state photon number on pump current (as well as clamping of the carrier density and gain above threshold). In the presence of dispersive loss, moderate nonlinearity ($\beta = -10^{-10}$) begins to modify the steady state behavior. For pump currents just above threshold, the behavior is close to linear. However, as the pump current increases, so does the loss, pulling down the input-output curve to a sub-linear behavior. For even stronger nonlinearity ($\beta = -10^{-9}$), a bistable transition occurs that creates a range of photon numbers which have no stable steady state solution. In particular, this occurs because there is a nonzero photon number at which the cavity experiences minimum loss. The topmost bistable branch (with $\kappa_n\gg 0$ and strongest squeezing) needs to be accessed hysteretically ``from above,'' by pumping to a high power (beyond the right bistable edge) and slowly lowering the power.

\subsection{Broadband intensity noise squeezing}
\label{sec:comparison}

We now describe how the mechanism of intensity-dependent loss can compress steady state photon statistics (Fig. \ref{fig:f3}a). The steady states of all lasers are characterized by a balance between saturable gain and loss. 
In a conventional laser with ``linear loss,'' the loss rate seen by the cavity field is the same for all photon numbers. For photon numbers where gain exceeds loss, an effective ``force'' encourages occupation of yet higher photon numbers; for photon numbers where loss dominates gain, an effective force encourages occupation of lower photon numbers. The intersection point where ``gain equals loss'' represents the equilibrium point between these two forces, and consequently determines the mean photon number of the cavity in the laser steady state. While the intersection point determines the mean photon number, the behavior of the photon number-dependent gain and loss in the vicinity of this intersection dictates the variance of the photon number probability distribution $p(n)$. In conventional lasers which are far above threshold, the probability distribution approaches that of a coherent state, with Poissonian statistics. 

This situation changes significantly when linear loss is replaced by a strongly intensity-dependent loss. If the loss rises sharply with photon number around its intersection with the saturable gain, then the steady state probability distribution becomes compressed compared to the case of linear loss. Intuitively, this is because the disparity between loss and gain around the steady state is magnified relative to the conventional laser, resulting in larger ``forces'' that squeeze the probability distribution to sub-Poissonian statistics. Roughly speaking, the photon number variance is determined by the ratio of the slopes of the gain and loss. This mechanism enables the sharp loss laser to create steady states with variance lower than the mean, a feature only possible in non-classical light. In the most extreme limit, the loss may rise so sharply that only a single number state (the mean) has a substantial probability of occupation, approaching a cavity Fock state. However, realizing intracavity Fock states would likely require systems with fewer photons and stronger nonlinearities, such as exciton-polariton condensates \cite{A:V1}.


To quantify this effect in semiconductor laser systems, we consider the photon number variance, given by $(\Delta n)^2=\frac{1}{\pi}\int_0^\infty d\omega \langle \delta n^\dagger(\omega) \delta n(\omega) \rangle$, where $\delta n(\omega)$ gives the spectrum of intensity fluctuations and is governed by Eq. \ref{eq:fluc}.
A useful parameter to quantify the quantum nature of light is the Fano factor, defined as $F=(\Delta n)^2/n$. The Fano factor is 1 for Poissonian light, corresponding to the shot noise limit; values below one indicate sub-Poissonian light below the shot noise level. We calculate the most general expression for $F$ (including carrier nonlinearity) in the presence of nonlinear dispersive loss in the S.I.. For weak Kerr and carrier nonlinearities, $F\rightarrow 1$ when pumping far above threshold, approaching Poissonian (coherent) statistics. Our main result here is that for strong Kerr nonlinearity ($n\kappa_n \gg \kappa_0,n|\kappa_N|,\gamma_\parallel,G_N$), the Fano factor behaves as
    \begin{align}
    F\rightarrow \kappa/(n\kappa_n)    
    \end{align}
    for large $n$. Critically, the ratio $\kappa_n/\kappa$ is a measure of how sharply the loss varies with $n$ compared to the absolute loss rate at the steady state photon number, and thus dictates the dimensionless ``sharpness'' of the loss. The Fano factor is inversely proportional to this sharpness factor, and thus sharp losses can lead to sub-Poissonian states.

In Fig. \ref{fig:f3}, we demonstrate the effects of intensity noise squeezing in semiconductor lasers with nonlinear dispersive loss. Just above the left point of bistability, $n$ stays approximately constant while the photon number variance $\Delta n^2$ can decrease sharply. In the plot of the Fano factor spectrum $\Delta n^2(\omega)/n$ (Fig. \ref{fig:f3}b), the intensity noise fluctuations associated with relaxation oscillations (ROs) are quenched closer to the left bistable point. Due to the sharp loss, the RO peak is in general significantly suppressed compared to the case of linear loss (the RO frequency and damping rate are increased in accordance with Eq. \ref{eq:ro}). Note that as a result of the bistability, the laser can exist in two states with very different photon numbers over a range of pump currents. The larger photon number branch corresponds to sharp loss ($\kappa_n>0$) in this scheme. Overall, nonlinear dispersive loss creates significant broadband intensity noise squeezing by orders of magnitude compared to analogous linear loss.

We also found that intensity noise squeezing can extend to the light which exits the cavity. To analyze this effect, the output noise spectrum can be computed from the intracavity noise spectrum by coupled mode theory (see S.I. for details). In Fig. \ref{fig:f3}c, we plot the output intensity spectrum normalized to the shot noise limit (SNL) for three different pump powers (with $\beta=-10^{-9}$). When a shot noise limited pump is used, output photon noise is not squeezed below the SNL. By using ``quiet'' pumping (i.e. constant current driving), it is possible to achieve noise reduction exceeding 10 dB below the SNL over GHz bandwidths. In conventional semiconductor lasers that are quietly pumped, weak output squeezing has been experimentally observed at sub-GHz bandwidths and strong squeezing ($>10$ dB) is only predicted to occur for operation far above threshold and is not associated with intracavity squeezing \cite{machida1987observation}. In contrast, the mechanism of nonlinear dispersive loss (1) creates strongly squeezed intracavity states, (2) strongly suppresses the relaxation oscillation peak ($>12$ dB relative to a conventional semiconductor laser with the same pump current but without nonlinear dispersive loss), extending output squeezing to GHz bandwidths, and (3) may allow significant output squeezing even at moderate pump currents owing to the bistability that creates strong intracavity squeezing near threshold for nonzero photon number.


Finally, we calculate noise frequency-integrated Fano factors as a function of pump current in Fig. \ref{fig:f3}e by integrating the spectra in Fig. \ref{fig:f3}b. For linear loss, the Fano factor approaches unity (shot noise limit) far above threshold. The behavior of Fano factor for nonlinear dispersive loss is phenomenologically different. For simplicity, in Fig. \ref{fig:f3}e, we only plot the sharp loss (upper) branch when bistability is present (purple curves). We note that the lower branch, accessible by normal pumping from threshold, resembles linear behavior apart from the bistable point, which creates a discontinuity in the Fano factor as a function of pump current. On the upper branch, linear behavior (shot noise) is restored when the detuning from the Fano resonance grows large ($\kappa_n\approx 0$). Approaching the left bistable edge, the cavity frequency approaches the Fano resonance and, for a certain $n$, the ratio $\kappa/(n\kappa_n)$ approaches a minimum, corresponding to maximum intracavity squeezing. The Fano factor does not decrease indefinitely due to intensity-carrier noise coupling and finite carrier noise from nonradiative decay processes. Nonetheless, low linear background losses, sharp dispersive dissipation, and large Kerr nonlinearities can create intracavity squeezing over 10 dB below the shot noise limit.

We next consider a second kind of dispersive loss -- distributed feedback provided by, for example, a distributed Bragg reflector (DBR). This type of loss marks a departure from the adiabaticity criterion that limits the sharpness of Fano-type losses because its timescale is instead set by the width of the DBR pass/stop band, not the sharpness of its decay. In principle, this means that the DBR-type loss can be made quite large, enhancing intensity noise squeezing further. This is shown in Fig. \ref{fig:f3}c,e, where sharper loss profiles (obtained by increasing the number of layers in the DBR) correspond to enhanced squeezing (5 dB lower than the Fano mirror example in Fig. \ref{fig:f3}d). Additionally, the sharp loss region ($\kappa_n>0$) in the case of DBR loss profiles can be accessed by pumping directly from threshold, where a stop band transitions to a pass band. Further details about the DBR example, including the exact analytical form for the loss, are provided in the S.I..

\subsection{IR and terahertz squeezing using quantum cascade lasers}
\label{sec:qcl}
\begin{figure}
    \centering
    \includegraphics[scale=0.65]{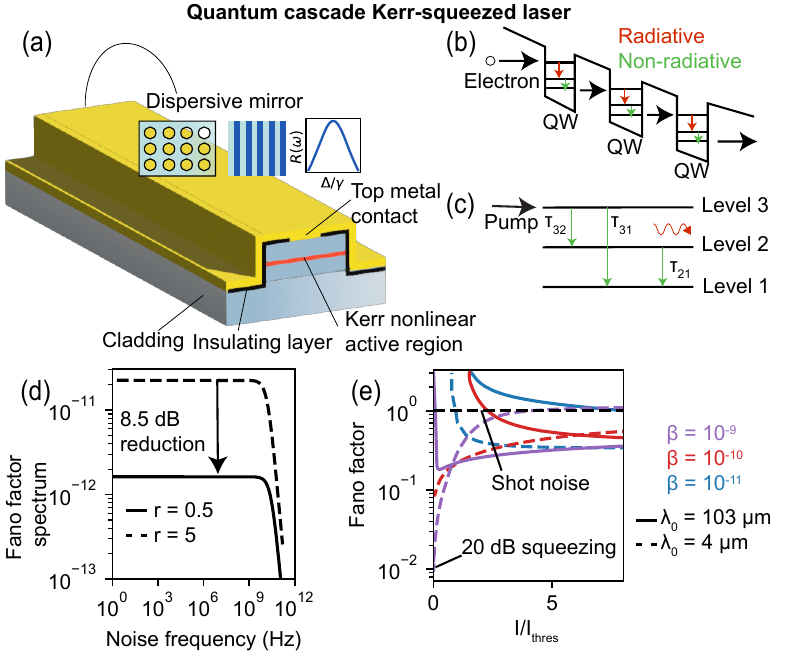}
    \caption{\textbf{Strongly squeezed IR and terahertz light using QCLs.} (a), (b) Basic dispersive Kerr-squeezed QCL laser architecture with nonlinear dispersive loss. Electrons make subband transitions in a given quantum well and tunnel to the next one. Dispersive outcoupling is provided by a photonic crystal fabricated on an end facet of the QCL. The giant, ultrafast Kerr nonlinearity of the active region due to intersubband transitions is used to generate nonlinear dispersive loss. (c) Three-level system used for rate equation analysis with nonradiative decay timescales from each level indicated. (d) Fano factor spectrum for two different pump strengths $r\equiv I/I_{\mathrm{thres}}$, with $\beta=10^{-9}$ and $\kappa_0=\mathrm{FSR}$. A similar bistability to the diode laser case is present here, and the $r=0.5$ curve is for the upper (low noise) branch ($r=5$ corresponds to large detuning from the Fano resonance and lies in the approximately linear loss regime). (e) Integrated Fano factor as a function of pump strength for the low noise branch for three different nonlinear strengths and operating wavelengths in the IR and THz. For these simulations, we use system parameters measured from experiment: wavelength $\lambda_0=4, 103$ $\mu$m (IR, THz), $\tau_{32}=2.1$ ps, $\tau_{31}=3.4$ ps, $\tau_{21}=0.5$ ps, $m=25$ gain stages, confinement factor $\Gamma=0.2$, cavity length $L=3$ mm, and gain coefficient $G_N=10^5\Gamma$ s\textsuperscript{-1} \cite{rana2002current,wang2018rate}. The Fano resonance has FWHM $\gamma=2\times10^{12}$ rad/s and is centered at $n_c=2.5\times 10^7$.}
    \label{fig:qcl}
\end{figure}

To emphasize the generality of the physics of nonlinear dissipation, we apply this mechanism to quantum cascade lasers (QCLs), showing that strong intensity squeezing can be extended to ``difficult'' spectral ranges where squeezing has not been demonstrated, such as the mid-IR and THz. QCLs employ intersubband transitions for stimulated emission, allowing recycling of the carrier population and therefore high output powers, since a single carrier can now generate $m$ photons if $m$ gain stages are used \cite{capasso2000new,bai2009intracavity}. This endows QCLs with giant intrinsic Kerr nonlinearities that have been employed in a variety of applications, such as frequency comb generation for molecular spectroscopy in the infrared \cite{villares2014dual}. We note that the picosecond timescale of these nonlinearities can fulfill the adiabaticity criterion for nonlinear dispersive loss \cite{weber2006theory} and that low-loss dispersive mirrors have been previously used to create dispersion-compensated QCL frequency combs \cite{lu2017dispersion}. Strongly intensity noise-squeezed light from QCLs, if realized, is extremely promising given that (1) intensity noise squeezing is more difficult to achieve in QCLs than other semiconductor lasers due to nonradiative decay of carriers in multiple levels \cite{rana2002current}, and (2) QCLs operate at wavelengths that are of great interest for sensing applications but are inaccessible by most other lasers. 

A sample design for a QCL with nonlinear dispersive loss is provided in Fig. \ref{fig:qcl}a. Here, the intrinsic Kerr nonlinearity of the active region combined with a dispersive mirror on the laser's output facet generates nonlinear dissipation. To quantify the steady state and noise behavior of this system, we proceed by a Langevin force-based rate equation analysis as before. We use a three-level model for the carrier dynamics (Fig. \ref{fig:qcl}b, c), with rate equations describing the evolution of the photon and carrier populations provided in the S.I.. The nonradiative decay time constants governing transitions between the three carrier levels are given by $\tau_{31},\tau_{32},\tau_{21}$ and linear gain proportional to the difference in population between levels 2 and 3 is assumed.

We calculate intracavity intensity noise spectra and integrated Fano factors by Fourier transforming the linearized rate equations (as done above), with details of the calculation provided in the S.I.. We find that the DC/low-frequency noise is suppressed by a factor $(n\kappa_n/\kappa)^2$ in the presence of strong nonlinear loss, $n\kappa_n\gg \kappa,1/\tau_{21},1/\tau_{31},1/\tau_{32}$. We plot the noise behavior for a sample system with Fano mirror outcoupling in Fig. \ref{fig:qcl}d, e. In this figure, we consider steady state and noise for three different Kerr nonlinear strengths $\beta$ and two different operating wavelengths $\lambda_0$ to mimic realistic experimental systems operating in the IR and THz. For comparison, we note that per-photon nonlinear strengths $\beta\sim 10^{-10}$ were observed nearly two decades ago when QCLs were first used for self-mode-locking \cite{paiella2000self}. Our results show that the strong, ultrafast Kerr nonlinearity in QCLs in combination with dispersive loss mechanisms can be harnessed to generate strong broadband intensity noise squeezing that has generally evaded mid-IR and THz wavelengths. Note also that (1) QCLs do not suffer from the GHz relaxation oscillations present in conventional semiconductor lasers due to the ultrafast (intersubband) carrier dynamics \cite{martini2001absence} and (2) linewidth enhancement due to free carriers is negligible, so the cavity resonance frequency $\omega_R(n,N)\rightarrow \omega_R(n)$ and the nonlinear loss $\kappa(n,N)\rightarrow \kappa(n).$ We finally note that self-pulsing by the mechanism in Sec. \ref{sec:mf} in QCLs may be possible but is more difficult to achieve than in conventional diode lasers. This is due to the ultrafast (intersubband) carrier dynamics in QCLs, which create high frequency (exceeding GHz) relaxation oscillations that are difficult to undamp.


\section{Discussion}
\label{sec:discussion}

We briefly describe some of the other experimental platforms for realizing the effects of nonlinear dispersive loss. We have already shown how quantum cascade lasers (QCLs) are promising realizations of semiconductor lasers with nonlinear dispersive loss given their giant, ultrafast Kerr nonlinearities. QCLs emit at IR and THz wavelengths, overlapping with the vibrational modes of many biochemically relevant molecules, making the possibility of developing quantum-enhanced chemosensors based on the principles described here tantalizing. 

Because semiconductor platforms are conducive to integration with on-chip photonic crystal optical elements, many designs have already achieved the dispersive losses considered here and therefore could exhibit intensity noise reduction if quality factors and nonlinear strengths are within the tolerances required. For example, previous work has realized ``Fano lasers'' that exhibit self-pulsing due to the interplay between dispersive loss and carrier nonlinearity \cite{yu2017demonstration}. A Fano resonance is created by coupling between a waveguide and nanocavity (point defect) in a photonic crystal slab. By increasing the quality factor of the lasing waveguide mode in these types of structures and integrating a Kerr material in/around the gain region, intensity noise reduction by nonlinear dispersive loss could be observable (Fig. \ref{fig:fock_schema}c).

Photonic crystal surface-emitting lasers (PCSELs) are another platform that may be used for demonstrating the effects of nonlinear dispersive loss \cite{noda2017photonic}. PCSELs may present advantages such as single-mode operation and high output powers; in contrast to the Fano laser concept, lasing occurs transversely (and thus the Fano mirror is aligned transversely rather than longitudinally). However, because losses may be significant in both longitudinal and transverse directions, it is necessary to optimize quality factors in both directions. The use of bound states in the continuum is also a promising way forward to achieve high quality factor resonances and nonlinear dissipation when PCSELs are endowed with strong Kerr nonlinearity \cite{hsu2016bound}.

In addition to Fano-type dispersive losses, distributed feedback-based losses have been commonly exploited to enforce single-mode operation. Examples include distributed Bragg reflector (DBR) fiber lasers, vertical cavity surface-emitting lasers (VCSELs), and DBR diode lasers \cite{carroll1998distributed, iga2018forty}. All of these architectures include sharply frequency-dependent elements that may be used to achieve strong noise condensation. High quality fabrication is necessary to minimize background losses (e.g., scattering) at interfaces in order to observe the intensity noise condensation described here.

Furthermore, we note that even stronger nonlinearities may be achievable in systems such as microdisk and quantum dot lasers due to enhanced confinement and ultralow mode volumes \cite{srinivasan2006cavity}. For mode volumes achieving $\lambda^3$, the dimensionless Kerr coefficient can be orders of magnitude larger than the values considered here. Lastly, combining our methodology with recent proposals for nanolasers with strong sub-wavelength confinement \cite{mork2020squeezing} could yield even further intensity noise reduction in output noise. 


\section{Outlook}
\label{sec:outlook}

In this paper, we have shown how semiconductor lasers with sharply frequency-dependent outcoupling and Kerr nonlinearity can be used to create lasers which possess intrinsic bistability and self-pulsing capabilities in the classical domain as well as high levels of quantum mechanical intensity noise squeezing both inside and outside the laser cavity. The squeezing occurs across a huge bandwidth in noise frequency, giving rise to near-Fock states with strong squeezing in photon number. Furthermore, we have shown that the squeezing is achievable from IR to THz wavelengths, potentially unlocking numerous applications in sensing, computing, and metrology. We anticipate that many existing experimental platforms could realize the intensity noise reduction, bistability, and self-pulsing effects described here, especially systems employing a geometry that maximizes photonic (Kerr) nonlinearity.

This work naturally suggests additionally possibilities for using nonlinear dispersive dissipation to control the output state of semiconductor lasers. Examples of topics for additional investigation include the effect of nonlinear dispersive loss on phase noise and linewidth, the effects of optical feedback on pulsing, bistability, and intensity/phase noise (e.g., in external cavity lasers), and the simultaneous control of self-pulsing and squeezing to generate pulsed squeezing. Semiconductor lasers are ubiquitous in many real-world applications and we envision that the use of nonlinear dispersive loss could render them novel tools to control the mean field and noise behavior of light across a wide range of wavelengths.

\section{Acknowledgements} We acknowledge useful discussions with Dima Kazakov and Yannick Salamin. S.P. acknowledges the financial support of the Hertz Fellowship Program, NSF Graduate Research Fellowship Program, and previous support of the Undergraduate Research Opportunities Program. J.S. is supported in part by a Mathworks Fellowship, and acknowledges previous support from a National Defense Science and Engineering Graduate (NDSEG) Fellowship (F-1730184536). N.R. acknowledges the support of a Junior Fellowship from the Harvard Society of Fellows, as well as earlier support from a Computational Science Graduate Fellowship of the Department of Energy (DE-FG02-97ER25308), and a Dean's Fellowship from the MIT School of Science. This material is based upon work supported in part by the Air Force Office of Scientific Research under the award number FA9550-20-1-0115; the work is also supported in part by the U. S. Army Research Office through the Institute for Soldier Nanotechnologies at MIT, under Collaborative Agreement Number W911NF-23-2-0121. We also acknowledge support of Parviz Tayebati.


\bibliographystyle{unsrt}
\bibliography{main.bib}

\setlength{\parindent}{0em}
\setlength{\parskip}{.5em}
\vspace*{-2em}

\end{document}


\rmfamily

\title{Supplementary information for: \\
Strong broadband intensity noise squeezing from near-infrared to terahertz frequencies in semiconductor lasers with nonlinear dissipation}
\author{Sahil Pontula$^{1,2,\dagger}$, Jamison Sloan$^{1,\dagger}$, Nicholas Rivera$^{1,3}$, and Marin Solja\v{c}i\'{c}$^{1,4}$}

\affiliation{$^{1}$Department of Physics, Massachusetts Institute of Technology, Cambridge, MA 02139, USA. \\
$^{2}$Department of Electrical Engineering and Computer Science, Massachusetts Institute of Technology, Cambridge, MA 02139, USA.  \\
$^{3}$Department of Physics, Harvard University, Cambridge, MA 02138, USA. \\
$^{4}$Research Laboratory of Electronics, Massachusetts Institute of Technology, Cambridge, MA 02139, USA. \\
$\dagger$ Denotes equal contribution. }
\noindent	

\noindent

\clearpage

\renewcommand{\sp}{\sigma_+}
\newcommand{\sm}{\sigma_-}

\setlength{\parindent}{0em}
\setlength{\parskip}{.5em}
\vspace*{-2em}


\newcommand{\bin}{b_{\text{in}}}
\newcommand{\bbarin}{\bar{b}_{\text{in}}}
\begin{abstract}
    In this Supplementary Information (S.I.), we present derivations of results reported in the main text, further details about the systems considered, and supplemental figures. Included in these discussions are closer examinations of mean-field self-pulsing and bistability, the effect of strong carrier nonlinearity on quantum noise, and a derivation of output noise in the presence of nonlinearity and dispersive loss.
\end{abstract}

\maketitle

\tableofcontents

\newpage

\section{Heisenberg-Langevin equations of motion}

The Hamiltonian of a simple two-band semiconductor can be writen as 
\begin{equation}
    H_{\mathrm{SC}} = \sum_q (\epsilon_g^{(0)} + \epsilon_{e,q}) c_q^\dagger c_q + \sum_q \epsilon_{h,q} h_q^\dagger h_q + V_{\text{int}}.
\end{equation}
Here, $c_q$ and $h_q$ are the fermionic annihilation operators for conduction-band electrons and valence-band holess at momentum $q$. They satisfy the fermionic commutation relations $\{c_q, c_{q'}^\dagger\} = \delta_{qq'}$, and likewise for $h_q$. Additionally, $\epsilon_g^{(0)}$ is the unrenormalized bandgap energy which separates the two bands at zero momentum. Also note that the sums $\sum_q$ are intended to note a sum over all electron states $q$, including momentum, spin, and anything else that might be relevant. Finally, $V_{\text{int}}$ represent interactions (collisions between electrons, interactions of the electron with the lattice, etc.). We will not need to consider the effects of this term, but its presence will lead to effects such as collision-induced equilibration of carriers within a band, relaxation of carriers from the upper band to the lower band, etc. Interactions can also lead to some shifts in the gain spectrum induced by carrier screening and band-gap renormalization.

Now, we would like to introduce a single mode Kerr nonlinear cavity which has frequency $\omega_0$ and annihilation operator $a$, so that the Hamiltonian of the cavity is $H_{\text{cavity}} = \omega_0 a^\dagger a \left(1+\beta a^\dagger a\right)$ with $\beta$ the per-photon Kerr nonlinearity. In order to describe lasing, the cavity should interact with the semiconductor gain medium through its dipole moment. We can define analogs of the atomic raising/lowering operators $\sigma_{\pm}$ for each electron label $q$ as $\sigma_q \equiv c_q h_{q}$. Then, in the rotating wave approximation (which assumes the light-matter coupling between the light and semiconductor is weak), the interaction between cavity and semiconductor is
\begin{equation}
    H_{\text{int}} = \sum_q \left(g_q a \sigma_q^\dagger + g_q^* \sigma_q a^\dagger \right).
\end{equation}
Then the Hamiltonian of the full laser is the sum of the contributions $H = H_{\text{SC}} + H_{\text{cavity}} + H_{\text{int}}$. Our goal then is to write equations of motion for quantities of interest, and then solve these equations for steady state, transient, and noise properties of the laser. To do so, we will now write Langevin equations of motion for the semiconductor laser. This amounts to computing the Heisenberg equations of motion for the operators of interest, adding the relevant pumping and damping terms, and finally computing the correlations between the Langevin forces which results to describe noise behavior.

For the polarization operator, we find
\begin{equation}
    \dot{\sigma}_q = -i\omega_q\sigma_q - \gamma_\perp \sigma_q + ig_q a(n_{e,q} + n_{h,q} -1) + f_q,
\end{equation}
where $\omega_q$ is the energy difference between the valence and conduction bands for state $q$. We see that $\sigma_q$ oscillates in the same way that $\sigma_i$ does for an atomic gain medium. Additionally, we see that the quantity in parentheses (which we shall define as $d_q$) in the second term acts like the inversion in an atomic gain medium. Specifically, the occupation operators for the electrons and holes can both takes values between 0 and 1. For a completely unexcited state ($n_e = n_h = 0$), the grouped quantity is $d_q = -1$. For a completely excited state ($n_e = n_h = 1$) we have $d_q = 1$. Thus $d_q$ can be thought of as the population inversion for each electron state $q$. 

For the cavity photon annihilation operator,
\begin{align}
\begin{split}
        \dot{a} &= -i\omega_0 \left(1+\beta a^\dagger a\right) a - \frac{\kappa}{2}a - i\sum_q g_q^* \sigma_q + f_a\\
        &= -i\omega_0 a \left(1+\beta a^\dagger a\right) a - \frac{\kappa}{2}a + \frac{a}{\gamma_\perp}\sum_q |g_q|^2 \mathcal{D}_qd_q + f_a,
\end{split}
\end{align}
where $\kappa$ is the cavity number/energy damping rate, $\mathcal{D}_q\equiv\frac{\gamma_\perp}{i(\omega-\omega_q)+\gamma_\perp}$ and $f_a$ is the Langevin force for the annihilation operator. In the second line, we adiabatically eliminated the polarization. Note that $\beta$ represents the per-photon Kerr nonlinear strength.

Lastly, for the electron occupation operator,
\begin{equation}
    \dot{n}_{e,q} = \Lambda_{e,q}(1 - n_{e,q}) - B_q n_{e,q} n_{h,q} - \gamma_{\parallel} n_{e,q} - \gamma_e(n_{e,q} - (n_{e,q})_0) + ig_q^* a^\dagger\sigma_q - ig_q \sigma_q^\dagger a + F_{e,q}.
\end{equation}

In order from left to right, the terms are
\begin{itemize}
    \item \textbf{Population pumping.} This is the pump rate due to carrier injection. When summing over this term, we get the actual pump rate $I$ at which free carriers are injected.
    \item \textbf{Loss due to spontaneous emission.} Excited carriers can be lost due to spontaneous emission. Since different $q$ can have different energy splittings, one of these spontaneous emission events will not necessarily be into the laser mode of interest. The coefficient $B_q$ is the rate for a particular momentum state $q$.
    \item \textbf{Nonradiative decay of population.} This term represents the rate at which exited carriers become unexcited in a manner which is proportional to the population (e.g., due to phonon emission).
    \item \textbf{Carrier-carrier relaxation.} This term represents relaxation to the equilibrium value $(n_{e,q})_0$ within a band. The fact that $\gamma_e$ tends to be very large compared to other relaxation rates allows one to make the so-called ``quasiequilibrium'' approximation in which each band acquires a Fermi-Dirac distribution. Moreover, because this term only redistributes carriers to different $q$ within the same band, it does not have an effect on the total inverted population. Thus, when summing this term over electron states, it vanishes.
    \item \textbf{Population depletion by stimulated emission into cavity mode.} This is the only term that can be derived from the Hamiltonian written above. This is the term that causes the population of excited states to deplete when stimulated emission occurs.
\end{itemize}

Now, we identify
\begin{align}
    G(N)(1-i\alpha_L) &\equiv \frac{2}{\gamma_\perp} \sum_q |g_q|^2 d_q \mathcal{D}_q \\
    \Gamma(N) &\equiv \gamma_\parallel N + \frac{1}{V}\sum_q B_q n_{e,q} n_{h,q} \\
    N &\equiv \sum_q n_{e,q}.
\end{align}
where the linewidth enhancement factor $\alpha_L\equiv\frac{d\chi_r/dN}{d\chi_i/dN}$, with $\chi=\chi_r+i\chi_i$ the susceptibility of the active material \cite{yamamoto1986amplitude}. We can now identify the resonance frequency using $a\equiv\alpha e^{i\phi}$ and $\dot\phi=\frac{1}{2}\frac{d}{dt}\ln\left(\frac{a}{a^\dagger}\right)=\frac{\dot a}{2a}-\frac{\dot a^\dagger}{2a^\dagger}$, showing that
\begin{equation}
    \omega_0 \rightarrow \omega_0\left(1+\beta n + \frac{\alpha_L}{2\omega_0} G(N)\right),
    \label{eq:om}
\end{equation}
so that the ``carrier nonlinearity'' is identified as $\sigma\equiv \alpha_L G_N/2\omega_0$. With these substitutions and neglecting the effects of spontaneous emission, the Heisenberg-Langevin equations in the main text are obtained.

Here, we neglected any frequency-dependent phase shifts imparted by the Fano mirror. These can be rigorously incorporated into the Heisenberg-Langevin equations using coupled mode theory, as we do below in Sec. \ref{sec:output}. The result is a phase shift $\tan^{-1}(\delta(\omega)/\gamma)$, where $\delta(\omega)$ represents the detuning from the Fano resonance and $2\gamma$ the width of the Fano resonance. We assume the second cavity mirror (back reflector) is broadband and imparts no phase shift. The effect of including the Fano mirror's phase shift is to make the resonance frequency no longer analytically solvable given $n,N$ using Eq. \ref{eq:om}. Instead, it must be solved numerically. However, we find that the effect of this dispersive phase shift is negligible over the detunings we consider: sweeping across the Fano resonance gives a deviation from the prediction of Eq. \ref{eq:om} of at most $0.02\gamma$, likely from the broad width we assume for the Fano resonance under the adiabatic approximation.

\subsection{Carrier equation of motion under optical excitation}

In the case of free carrier excitation due to optical pumping, the mean field carrier equation of motion derived from the Heisenberg-Langevin formalism reads \cite{yu2017demonstration}
\begin{align}
    \dot N = \frac{\eta P_p}{\hbar\omega_p V_p} - \gamma_\parallel N - nG(N), 
\end{align}
where $\eta$ is the pump efficiency, $P_p$ the pump power, $\hbar\omega_p$ the energy of a pump photon, and $V_p$ the pump volume. Assuming a pump volume $V_p$ on the order of the active region volume and excitation by a near-IR source (around 800 nm), typical pump powers are on the order of tens of mW for examples considered in the main text with pump currents on the order of tens of mA.

\subsection{Estimation of per-photon Kerr nonlinearity $\beta$}

We briefly describe how the per-photon Kerr nonlinearity $\beta$ can be estimated. Previous work has derived the per-photon Kerr nonlinearity from a quantum mechanical Hamiltonian approach \cite{drummond1980quantum}:
\begin{equation}
    \beta = \frac{3\hbar\omega_0}{8\epsilon_0^2}\int \chi^{(3)}(\textbf{r})|\textbf{u}(\textbf{r})|^4 d^3\textbf{r},
\end{equation}
where the electric field profile is normalized as $\int|\textbf{u}(\textbf{r})|^2\epsilon_r(\textbf{r}) d^3\textbf{r}=1$. To get an estimate of achievable $\beta$, we consider a buried heterostructure laser with GaAs gain and Al\textsubscript{0.34}Ga\textsubscript{0.66}As cladding. The active region has dimensions $0.1$ $\mu$m $\times$ $5$ $\mu$m $\times$ $1$ mm, the lasing frequency is near the bandgap of GaAs, $\omega_0=2.16\times 10^{15}$ rad/s, and the refractive indices of GaAs and Al\textsubscript{0.34}Ga\textsubscript{0.66}As are 3.6051 and 3.3734 respectively. We take $n_2\approx -10^{-16}$ m\textsuperscript{2}/W for Al\textsubscript{0.34}Ga\textsubscript{0.66}As \cite{boyd2020nonlinear}. We solve Maxwell's equations in the core and cladding using a slab waveguide model, obtaining a confinement factor $\Gamma\approx 0.3$ and per-photon Kerr nonlinearity $\beta\approx -6\times 10^{-10}$. 




\section{Mean-field dynamics: bistability and self-pulsing}


\subsection{Bistability due to Kerr nonlinearity}

Here, we quantify the bistability boundaries that arise when intensity-dependent loss is present. As shown in Fig. 2, this bistability correlates with the phenomenon of self-pulsing and demarcates an unstable region in the S-curve for the laser. Its boundaries can be found by noting that, in the steady state,
\begin{align}
    \begin{split}
        I(n)&=\gamma_\parallel N(n) + nG_N\left(N(n)-N_{\mathrm{trans}}\right)\\
        N(n)&=\frac{\kappa(n)}{G_N}+N_{\mathrm{trans}}.
    \end{split}
\end{align}

The bistability boundaries (in pump $I$) are those values $I(n)$ for which $dI/dn=0$, for which we require
\begin{equation}
    \frac{dI}{dn}=0 \implies\kappa_n\left(\frac{\gamma_\parallel}{G_N}+n\right)+\kappa(n)=0.
\end{equation}
One can see, for example, that in the absence of intrinsic loss, $n_c$, the point of zero loss, satisfies this condition, since $\kappa_n(n_c)=\kappa(n_c)=0$.

\subsection{Onset and cessation of self-pulsing}

Self-pulsations begin when relaxation oscillations become undamped, $\Gamma_1<0$ and $\Omega^2_R>0$ (in the initial steady state solution). They do not, however, persist throughout the entire region where $\Gamma_1<0$, as shown in Fig. 2 of the main text. When the laser begins at the left edge of bistability in the S-curve (as in Fig. 2b) at steady state, the pulsations are transient and eventually collapse to the steady state solution at the center of the Fano resonance with minimum loss at photon number $n_c$ (this is the leftmost point of bistability). Note that the laser began at the second photon number $n^*$ (low intensity branch) that corresponds to the same pump power as photon number $n_c$. Eventually, we sweep through initial steady state photon numbers within the region of instability (still within bistable operation) that is not normally accessible by pumping directly from threshold. When the right edge of bistability is finally crossed, the laser enters the region with $\kappa_n>0$, characterized by heavily damped relaxation oscillations and intensity noise squeezing. When pumping from threshold, the laser jumps from the low intensity to high intensity branch at the right bistable edge.

\subsection{Pulse characteristics in self-pulsing regime}

\begin{figure}
    \centering
    \includegraphics{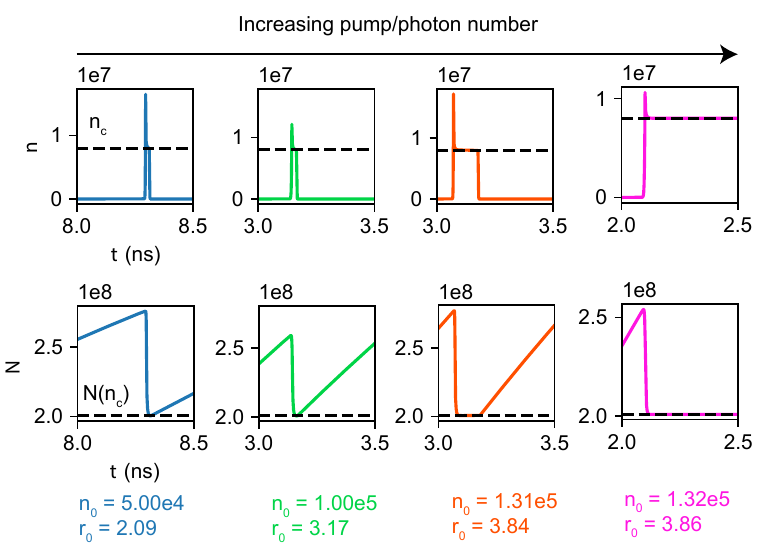}
    \caption{\textbf{Evolution of the pulse profile for carrier density and photon number from the self-pulsing to collapsed pulse regimes.} As the initial photon number $n_0$ approaches the left bistable edge, the pulse plateaus for longer at the center of the Fano resonance. Thus, the effective width of the pulse is dynamic within the regime over which self-pulsing occurs, depending on the initial state's proximity to the left bistable edge. Once the left bistable edge is crossed, the pulse collapses to a CW solution at higher photon number than the initial state. Here, $r_0$ denotes the initial pumping rate relative to threshold.}
    \label{fig:pw}
\end{figure}

The self-pulsations demonstrate an interesting behavior in pulse shape, as shown in Fig. \ref{fig:pw}. The initial sharp rise in the pulse profile is due to the undamping of relaxation oscillations, and its timescale is thus set by $1/|\Gamma_1|_{\mathrm{max}}\approx 1/|n\kappa_n|_{\mathrm{max}}$ ($\mathcal{O}(1)$ ps in our simulations). The same timescale characterizes the final drop in pulse power. In between these two features, two further timescales are at play. The decay after peak pulse power is initially very fast ($\mathcal{O}(1)$ ps) due to the strong damping of relaxation oscillations in the $\kappa_n>0$ region. The decay slows as the photon number approaches $n_c$, governed by $\Gamma_1$ evaluated at $n\approx n_c$. The final feature also sets the longest timescale for the pulse. It is a plateau near $n\approx n_c$ that emerges from ``quasi'' steady state conditions. The carrier density can be calculated by solving the carrier equation of motion in the steady state as 
\begin{align}
    N\approx \frac{I_0+n_cG_NN_{\mathrm{trans}}}{\gamma_\parallel + n_cG_N},
\end{align}
where $I_0$ denotes the (fixed) pumping rate. Notice here that for $n<n^*$, $N<N(n_c)$, so that $G(N)<\kappa(n)$ as the photon number drops below $n_c$ and approaches the point where the pulsing continues ($\Gamma_1<0$). The timescale for the plateau is then given by $\tau_p=1/|G(N)-\kappa(n_c)|$. Notice that $N\rightarrow N_c,\tau_p\rightarrow\infty$ as the initial steady state photon number $n_0\rightarrow n^*$. When $n_0=n^*$, the pulsations are transient and the laser approaches a steady state at $n_0=n_c$, as shown in Fig. \ref{fig:pw}. For $n_0\approx n^*$, the plateau changes based on the initial steady state (i.e. pumping rate) and can approach timescales of tens to hundreds of ps. 

The peak pulse power is more difficult to predict, depending on the initial fluctuation from steady state. However, it must occur at $n>n_c$ to saturate the pulse and begin its decline.

The pulse repetition rate is set by the carrier density recovery timescale when the pulse is off. During this time, the photon number $n\approx 0$, so that the mean-field dynamics of carrier density are given by
\begin{align}
    N(t)=\left(N_{\mathrm{min}}-\frac{I}{\gamma_\parallel}\right)e^{-\gamma_\parallel t} + \frac{I}{\gamma_\parallel},
\end{align}
where $I$ denotes the pump current and $N_{\mathrm{min}}$ the minimum carrier density. If $\Delta N=N_{\mathrm{max}}-N_{\mathrm{min}}$ is the difference in carrier density at the pulse maximum and minimum, the period between pulses is given roughly by
\begin{align}
    T_{\mathrm{rep}}\approx \frac{1}{\gamma_\parallel}\ln\left(\frac{I/\gamma_\parallel-N_{\mathrm{min}}}{I/\gamma_\parallel-N_{\mathrm{max}}}\right).
\end{align}
For the system parameters in the main text, $T_{\mathrm{rep}}\sim 10$ ns (100 MHz repetition rate). 

\section{Intensity noise}

\subsection{Langevin force correlators}

In this section, we derive the photon number correlator in the presence of two-photon absorption (TPA). We begin with the equation of motion for photon number probabilities in the presence of TPA only, $\dot p_n=-\frac{\alpha_{\mathrm{TPA}}}{2} n(n-1)p_n + \frac{\alpha_{\mathrm{TPA}}}{2} (n+1)(n+2) p_{n+2}$, where $p_n$ denotes the probability of having $n$ photons inside the laser cavity. Thus,

\begin{equation}
\begin{split}
        \langle \dot n\rangle &= \sum_j j\dot p_j \\ &= -2\alpha_{\mathrm{TPA}} \sum_j j(j-1)p_j \\ &= -\alpha_{\mathrm{TPA}} [\langle n^2\rangle - \langle n \rangle]. 
\end{split}
\end{equation}

\noindent The RHS reduces to $-\alpha_{\mathrm{TPA}}\langle n\rangle ^2$ assuming mean field theory, $\Delta n\ll \langle n\rangle,$ recovering the equation of motion $\dot n = -\alpha_{\mathrm{TPA}}n^2$. Using the generalized Einstein relation, the correlator is $\langle  2D_{nn}\rangle = \frac{d}{dt}\langle n^2 \rangle - 2\langle nD_n\rangle$, where we express $\dot n = D_n+F_n,$ with $D_n$ a diffusion term and $F_n$ a Langevin force. Thus 

\begin{equation}
\begin{split}
        \langle  2D_{nn}\rangle &= \left(\sum n^2 \dot p_n\right) + 2\alpha_{\mathrm{TPA}}\langle n^3-n^2\rangle \\
    &= -\alpha_{\mathrm{TPA}}\langle n(n-1)^2\rangle + 2\alpha_{\mathrm{TPA}} \langle n^2(n-1)\rangle \\
    &\approx 2\alpha_{\mathrm{TPA}} \langle n\rangle ^2,
\end{split}
\end{equation}

\noindent again assuming mean field theory. Allowing for one-photon gain and loss, $\langle  2D_{nn}\rangle=2\kappa n+\alpha_{\mathrm{TPA}} n^2$. The other nonzero diffusion coefficients are $\langle F_N^\dagger F_N\rangle = \langle 2D_{NN} \rangle= \epsilon I + R_{sp}n+\gamma_\parallel N, \langle F_N^\dagger F_n\rangle = \langle 2D_{Nn} \rangle = -Rn, \langle F_\phi^\dagger F_\phi\rangle = \langle 2D_{\phi\phi} \rangle = R_{sp}/2n$ where $R_{sp}\approx G(n,N)$ denotes the rate of spontaneous emission into the cavity mode, $R_{abs}\approx 0$ denotes the rate of absorption (negligible above threshold), $R=R_{sp}+R_{abs}$, and $\epsilon=0$ (1) for quiet (noisy) pumping. These correlators can be derived by computing the Einstein diffusion coefficients \cite{chow2012semiconductor} and give rise to nonzero fluctuations in $n,N$ about their steady state values. For intracavity noise calculations in the main text, pump noise is always included. Output noise calculations are performed for both noisy and quiet pumping schemes.

\subsection{Analytic intensity noise spectra and Fano factor expressions}
\label{sec:deriv}

In this section, we provide a linearization of the semiconductor laser rate equations in the presence of various nonlinearities and calculate relative intensity noise using this formalism. We obtain

\begin{equation}
\begin{split}
    \delta \dot n  &= -\left(\kappa_n n + \frac{pG_0}{2(1+p)} \right) \delta n + n\left(G_N-\kappa_N\right) \delta N + F_n-n\kappa_\omega F_\phi \\
    \delta \dot N &= -\left(\frac{G_0(1+p/2)}{1+p}-I_n\right) \delta n  - \left(G_Nn + \gamma_\parallel\right)\delta N + F_N.
    \end{split}
    \label{eq:full_linearized}
\end{equation}

\noindent where $p=n/n_{\mathrm{sat}}$ denotes the saturation fraction for photon number and $I_n\equiv dI/dn$ denotes carrier generation by TPA. Note that $G_0,G_N$ implicitly include the effects of gain saturation, $G_{0,N}\rightarrow G_{0,N}/\sqrt{1+p}$. Results in the main text assume $p,I_n\rightarrow 0$. 


For simplicity of notation, we will introduce $a=nG_N+\gamma_\parallel, b=n\left(G_N-\kappa_N\right), c=G_0\frac{1+p/2}{1+p} - I_n, d=n(\kappa_n-G_n), \Gamma_1 = a+d, \Omega_R^2 = ad+bc$. Note that $\Omega_R^2$ denotes the approximate relaxation oscillation frequency and $\Gamma_1$ the decay of relaxation oscillations. Fourier transforming the linearized rate equations,

\begin{equation}
    \begin{bmatrix}
-i\Omega + d & -b\\
c & -i\Omega + a
\end{bmatrix}
    \begin{bmatrix}
\delta n(\Omega)\\
\delta N(\Omega)
\end{bmatrix}
=
\begin{bmatrix}
F_n-n\kappa_\omega F_\phi\\
F_N
\end{bmatrix},
\end{equation}

yielding 
\begin{equation}
    \begin{bmatrix}
\delta n(\Omega)\\
\delta N(\Omega)
\end{bmatrix}
=\frac{1}{-\Omega^2+(ad+bc)-i\Omega(a+d)}    
\begin{bmatrix}
(-i\Omega+a)\left(F_n-n\kappa_\omega F_\phi\right)+bF_N\\
-c\left(F_n-n\kappa_\omega F_\phi\right)+(-i\Omega + d)F_N.
\end{bmatrix}
\end{equation}

The intensity noise spectrum is then

\begin{equation}
    \langle \delta n^\dagger(\Omega) \delta n(\Omega) \rangle = \frac{(\Omega^2+a^2)[\langle 2D_{nn} \rangle + n^2\kappa_\omega^2\langle 2D_{\phi\phi} \rangle]  + b^2\langle 2D_{NN} \rangle+ 2ab\langle 2D_{Nn} \rangle}{(\Omega^2-\Omega_R^2)^2+\Omega^2\Gamma_1^2}.
    \label{eq:rin}
\end{equation}

As a side note, ignoring the effect of Kerr nonlinearity but including dispersive loss and the associated amplitude-phase coupling, we see that RIN can be reduced by a factor $(1+\kappa_{\omega}^2)/(1-\alpha_L\kappa_{\omega}/2)^2\rightarrow 1/(1+\alpha_L^2)$ if the slope $\kappa_{\omega}$ is chosen appropriately, in agreement with earlier work on amplitude-phase decorrelation (where intensity noise is reduced somewhat at the expense of an increase in phase noise) \cite{newkirk1991amplitude}. However, this method leads to frequency selective squeezing, as opposed to the type of broadband squeezing we consider here. 

We compute the Fano factor from Eq. \ref{eq:rin} using the integrals 

\begin{align*}
    I_1&= \int_0^\infty \frac{1}{(\omega^2-x^2)^2+y^2}d\omega=\frac{\pi}{4y}\frac{\sqrt{2x^2+2\sqrt{x^4+y^2}}}{\sqrt{x^4+y^2}} \\
            I_2&=\int_0^\infty \frac{\omega^2}{(\omega^2-x^2)^2+y^2}d\omega=
    \frac{\pi}{4}\frac{\sqrt{-2x^2+2\sqrt{x^4+y^2}}}{\sqrt{x^4+y^2}}+x^2I_1,
\end{align*}

\noindent where $x,y\in \mathbb{R}.$ With $x^2=\Omega_R^2-\frac{\Gamma_1^2}{2}, y^2 = \Gamma_1^2\left(\Omega_R^2-\frac{\Gamma_1^2}{4}\right)$, we have $I_1=\frac{\pi}{2\Gamma_1\Omega_R^2}, I_2=\frac{\pi}{2\Gamma_1}$. Thus, the Fano factor reads

\begin{equation}
        F=\frac{1}{2n\Gamma_1\Omega_R^2}\left([\langle 2D_{nn} \rangle + n^2\kappa_{\dot\phi}^2 \langle 2D_{\phi\phi} \rangle] (\Omega_R^2+a^2) +\langle 2D_{Nn}\rangle ab  + \langle 2D_{NN}\rangle b^2\right)
\end{equation}

We now consider limiting expressions for $F$ in various limiting cases:

\begin{enumerate}
    \item For weak Kerr and carrier nonlinearities, $\kappa_n,\kappa_N\rightarrow 0$, we have $F\rightarrow 1+\kappa/(nG_N)$ when pumping far above threshold, recovering linear behavior. When $n$ becomes large far about threshold, the Fano factor approaches 1, resulting in Poissonian (coherent) statistics.
    \item For strong Kerr nonlinearity but weak carrier nonlinearity, $n\kappa_n \gg \kappa_0,n|\kappa_N|,\gamma_\parallel,G_N$, the Fano factor $F\rightarrow \kappa/(n\kappa_n)$ for large $n$, resulting in squeezing when $n\kappa_n>\kappa$.
     \item For strong carrier nonlinearity but weak Kerr nonlinearity, $n|\kappa_N| \gg \kappa_0,n|\kappa_n|,\gamma_\parallel$, we have $F\rightarrow \kappa/(nG_N)+G_N/|G_N-\kappa_N|\rightarrow G_N/|G_N-\kappa_N|$ for large $n$. The carrier nonlinearity reduces dependence of the rate of change of intensity fluctuations on carrier fluctuations ($G_N\rightarrow G_N-\kappa_N$), lowering the relaxation oscillation frequency $\Omega_R^2$ while leaving the damping of these oscillations unchanged. This can amplify low-frequency intensity noise slightly.
    \item For simultaneously strong Kerr and carrier nonlinearities, $n|\kappa_{n,N}| \gg \kappa_0,\gamma_\parallel$,
    \begin{equation}
    \begin{split}
        F&\rightarrow \frac{\kappa}{n|G_N+\kappa_n|}\left(1 + \frac{nG_N^2}{n\kappa_n G_N+|G_N-\kappa_N|\kappa}\right).
    \end{split}
        \label{eq:ffactor}
    \end{equation}
    Roughly, this expression can be broken into Kerr nonlinearity (first term) and carrier nonlinearity (second term) contributions. The former describes squeezing via increased $\Omega_R^2$ and damping of relaxation oscillations due to ``sharp'' intensity-dependent loss, while the latter reduces intensity noise-carrier noise coupling and thus $\Omega_R^2$. Kerr and carrier nonlinearities may therefore have competing effects, leading to interesting steady state and noise fluctuation behavior. 
\end{enumerate}
\subsection{Noise reduction using two photon absorption (TPA)}

Two photon absorption (TPA), though not a dispersive loss, is weakly nonlinear in photon number and thus may be expected to permit some squeezing in intensity noise. When TPA is present, for large $n$,

\begin{equation}
F\rightarrow \frac{3\kappa}{2n(G_N+\alpha_{\mathrm{TPA}})}\left(1+\frac{G_N}{\alpha_{\mathrm{TPA}}+\kappa/n}\right),    
\end{equation}

where $\alpha_{\mathrm{TPA}}=\kappa_n$. The minimum achievable Fano factor is 3/4, obtained when $\kappa_0/n\ll \alpha_{\mathrm{TPA}}\ll G_N$ (here $\kappa_0$ denotes linear background loss). To obtain the TPA coefficient $\alpha_{\mathrm{TPA}}$, we use the relationship between intensity $I$ and photon number $I\sim n\hbar\omega c/V$, so that $\alpha_{\mathrm{TPA}}\sim 2\hbar\omega cL\beta_{\mathrm{TPA}}\cdot \mathrm{FSR}/V$, where $L,V$ respectively denote the length and volume of the cavity. For a cavity field oscillating at $\omega\sim 10^{15}$ Hz for GaAs at 1064 nm ($\beta_{\mathrm{TPA}}=260 \text{ m/TW}$), we find $\alpha_{\mathrm{TPA}}\sim 10^{-8} \cdot \mathrm{FSR}$ for $L\approx 1$ mm, $V\approx 10^{-16}$ m\textsuperscript{3}. For typical intracavity photon numbers, the TPA contribution to the loss is then $10^{-2} \cdot \mathrm{FSR}$, a weak nonlinear background loss that is neglected for the examples in the main text where the primary nonlinear dispersive loss is much stronger.

\begin{figure}
    \centering
    \includegraphics{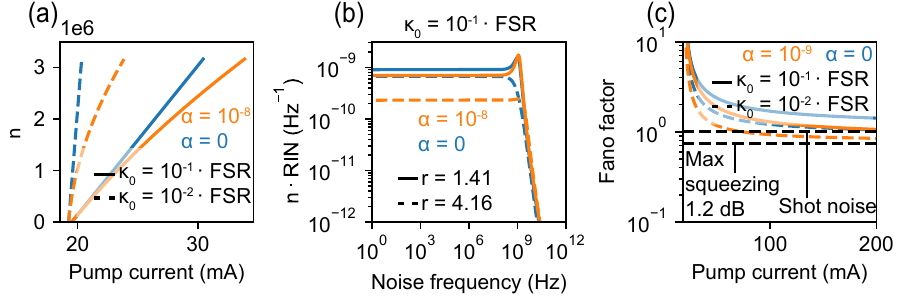}
    \caption{\textbf{Steady state and noise plots for two photon absorption.} (a) Steady state intracavity photon number as a function of pump current (S-curve), demonstrating sub-linear dependence of photon number with pump current for two-photon absorption (TPA). (b) Photon number variance spectrum for two different pump powers $r=I/I_{\mathrm{thres}}$, with broadband squeezing for intensity-dependent TPA. (c) Fano factor plots for linear and TPA loss profiles. The intensity dependence of TPA $\kappa(n)\propto n$ creates small ($<2$ dB) drops in Fano factor below the shot noise limit when pumped far above threshold. Here, $\alpha\equiv \alpha_{\mathrm{TPA}}/\mathrm{FSR}$.}
    \label{fig:tpa}
\end{figure}

As shown in Fig. \ref{fig:tpa}a, TPA creates a sublinear S-curve that arises from the monotonic dependence of $\kappa(n)$ on $n$. Fig. \ref{fig:tpa}b demonstrates how TPA induces broadband intensity noise squeezing, resulting in a weak suppression of Fano factor (integrated over all noise frequencies) in Fig. \ref{fig:tpa}c. Linear loss asymptotes to unit Fano factor for large pump powers, while TPA can result in minor noise condensation (though this effect can be washed out if TPA is too strong or too weak, in violation of $\kappa_0/n\ll \alpha_{\mathrm{TPA}}\ll G_N$). The source of Fano factor reduction for higher pump currents is slightly different for both loss profiles. For linear loss, it occurs because steady state $n$ increases linearly with pump current while the fluctuations $(\Delta n)^2$ have a sublinear dependence on pump current. In contrast, for TPA, the photon number $n$ is clamped at high pump current and the photon number distribution is squeezed slightly due to the nonlinear loss $\kappa(n)$.

\subsection{Noise reduction using nonlinear distributed feedback-based loss}

In this section, we consider distributed feedback semiconductor lasers where a distributed Bragg reflector (DBR) is fabricated on one (or both) ends/facets of the laser cavity, or a VCSEL-type structure is employed. In this case, we use the analytical form for DBR reflectivity given by coupled mode theory \cite{brovelli1995simple, yen1976gaas} to obtain

\begin{equation}
        \kappa(\omega)=-\mathrm{FSR}\cdot\log\left\vert \frac{g\sinh(\theta)}{\Gamma\cosh(\theta)+(\alpha_{\mathrm{DBR}}+i\delta)\sinh(\theta)}\right\vert^2,
        \label{eq:dbr_loss}
\end{equation}

where $\beta=\omega\tilde{n}/c$ is the propagation constant (wavevector), $g=\omega\Delta n/(\pi c)$ is the approximate coupling coefficient, $\delta=\beta-\pi/d$, $\Gamma^2=g^2+(\alpha_{\mathrm{DBR}}+i\delta)^2$, $\theta=N_{\mathrm{DBR}}d\Gamma$, and $\alpha_{\mathrm{DBR}}$ the radiative loss from the DBR. Here, $N_{\mathrm{DBR}}$ denotes the number of pairs of layers in the DBR, $d$ the thickness of a pair of layers, $\Delta n$ the index contrast, $\tilde n$ the effective index, and $\omega\equiv \omega(n,N)$ the laser frequency. Note that $\delta$ has the interpretation of a detuning from the Bragg value $\pi/d$ (the center of the Bragg stop-band of maximum reflectivity and thus lowest loss is at $\delta=0$). We would like to operate in the ``sharp loss'' regime, which is where the stop-band switches over to a pass-band, first occurring when $\theta=\pi\implies \delta^2 - g^2=\pi^2/L^2$. For a lossless DBR, choosing the frequency $\omega_c$ at which this sharp transition occurs fixes $\delta$ and therefore $\Delta n$ from the above relations:

\begin{equation}
    \Delta n = \frac{\pi c}{\omega_c}\sqrt{\left(\frac{\tilde{n}}{c}(\omega_c-\omega_t)\right)^2-\left(\frac{\pi}{L}\right)^2},
    \label{eq:index}
\end{equation}

\noindent where $\omega_t$ denotes the center of the stop band, so that $\omega_c-\omega_t$ is effectively the half-width of the stop band. The coupling coefficient $g$, index contrast $\Delta n$ and stop band width $2(\omega_t-\omega_c)$ are thus closely related.

\begin{figure}
    \centering
    \includegraphics[scale=0.5]{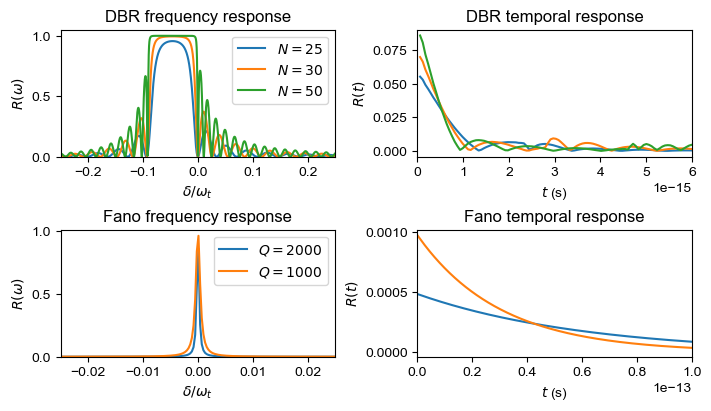}
    \caption{\textbf{Comparison of frequency and temporal response for Fano mirror and DBR losses.} In the top row, the DBR stop band is made sharper and its width is increased by increasing the number of layers. This results in a shorter, ultrafast relaxation time. In contrast, in the bottom row, the Fano mirror frequency response is made sharper by increasing the quality factor ($Q$) of the resonance, which has the effect of decreasing the width of the resonance while increasing its lifetime.}
    \label{fig:dbr_fano}
\end{figure}

To use Eq. \ref{eq:dbr_loss}, it is necessary to ensure the time response of the DBR is much faster than the free spectral range. We extract this time response by performing an FFT of $R(\omega)$. For lossy DBRs, $R(\omega)$ approaches a Lorentzian with width governed by $\alpha_{\mathrm{DBR}}$, and the maximum reflectivity may be far from unity. When the DBR is lossless, an analytical expression for the time response is in general difficult to obtain. We observe that the time response is faster for DBRs of larger bandgap (wider stop bands). Intuitively, outcoupling in a lossless DBR is through the coupling coefficient $g$ which scales with the index contrast $\Delta n$ and thus correspondingly with the stop band width $2(\omega_c-\omega_t)$. This is distinct from the Fano resonances considered in the main text where the loss profile was derived from interference between a ``direct channel'' pathway bypassing the Fano resonance and an ``indirect pathway'' coupling to an intrinsic resonant mode of the photonic crystal. In such a case, the time response of the effective nonlinear dispersive loss is governed by the the complex resonance frequency of the Fano resonance (intuitively, how long light spends trapped in the photonic crystal). Here, however, sharply frequency-dependent loss arises from a different mechanism, namely the photonic bandgap of the DBR. A comparison of the two different types of temporal responses are provided in Fig. \ref{fig:dbr_fano}.

The sharpness of $\kappa(n)$ increases with the number of layers $N_{\mathrm{DBR}}$ and Kerr nonlinear strength (the former corresponds to sharper evanescent decay of modes in the photonic bandgap). For the strongest nonlinearity and sharpest $R(\omega)$, multiple stop bands may be accessible, corresponding to multiple regions of noise condensation and bistability for each transition from pass band to stop band. An important distinction from the Fano resonances considered earlier is that the sharp loss regions $\kappa_n>0$ are now the lower bistability branches, accessible by simply pumping smoothly from threshold.

From an experimental standpoint, the sharpest loss (and strongest squeezing) can be obtained by maximizing the stop band width and number of layer pairs $N_{\mathrm{DBR}}$. The former is limited by the intracavity saturation intensity and required index contrast $\Delta n$, while the latter is limited by fabrication methods. Nevertheless, carefully-engineered DBR-based losses when coupled to strong Kerr nonlinearity in semiconductor lasers may result in unprecedented broadband intensity noise squeezing.

\section{Output photon noise}
\label{sec:output}

Here, we develop formalism to compute output photon noise in the presence of nonlinearity and dispersive dissipation. Let $a, d$ respectively denote the nonlinear cavity mode and the Fano mirror mode, both of which couple to a continuum of far-field modes $s_k$ (here $k$ labels momentum). From the full Hamiltonian of the system \cite{A:V1}, the Heisenberg equations of motion can be derived as
\begin{align}
    \begin{split}
        \dot a &= -i\omega_a(1+\beta a^\dagger a) a + G(N)(1-i\alpha_L)a - i\sum_k g_k^* s_k + F_G \\
        \dot d &= -i\omega_d d - i\sum_k v_k^* s_k \\
        \dot s_k &= -i\omega_k s_k - i (g_ka + v_k d).
    \end{split}
\end{align}
where $\omega_{a,d}$ denote the resonance frequencies of the cavity and Fano mirror, $\beta$ is the per-photon Kerr nonlinearity, $G(N)$ is the carrier-dependent gain (added phenomenologically), $\alpha_L$ is the linewidth enhancement factor, and $g_k,v_k$ are the couplings of $a,d$ to the far-field mode $s_k$. $F_G(t)$ is a Langevin force term for the gain. We neglect direct coupling between $a$ 
and $d$, though this can be readily incorporated into the Heisenberg equations. We can solve for $s_k$ as
\begin{align}
    s_k(t) = s_k(0) e^{-i\omega_k t} - i\int ^t dt' e^{-i\omega_k(t-t')}[g_ka(t')+v_kd(t')].
\end{align}
We assume momentum independent coupling $g=g_k,v=v_k$ and introduce $\kappa_\mathrm{FSR}=\pi \rho|g|^2,\gamma=\pi \rho |v|^2$. The input-output relation can then be found by taking $t\rightarrow \infty$, performing an integral over $k$, and Fourier transforming: 
\begin{align}
    s_\mathrm{out}(\omega) = -s_\mathrm{in}(\omega) + \sqrt{2\kappa_\mathrm{FSR}}a(\omega) + \sqrt{2\gamma}d(\omega).
    \label{eq:io1}
\end{align}
 We can write the Fourier transformed Heisenberg equations for $a,d$ as
\begin{align}
    \begin{split}
        -i\omega a &= -i\omega_a(1+\beta a^\dagger a) a + [G(N)(1-i\alpha_L)-\kappa_\mathrm{FSR}]a - \sqrt{\kappa_\mathrm{FSR}\gamma}d + \sqrt{2\kappa_\mathrm{FSR}}s_\mathrm{in} + F_G \\
        -i\omega d &= -i\omega_d d - \gamma d - \sqrt{\kappa_\mathrm{FSR}\gamma}a + \sqrt{2\gamma}s_\mathrm{in}. \\ 
    \end{split}
\end{align}
Eliminating $d(\omega)$ as
\begin{align}
    d(\omega) = \frac{\sqrt{2\gamma}s_\mathrm{in}-\sqrt{\kappa_\mathrm{FSR}\gamma}a}{i\delta_d+\gamma},
\end{align} 
with $\delta_d\equiv \omega_d-\omega$, we can write the Fourier transformed equation of motion for $a$ and the input-output relation as
\begin{align}
    -i\omega a &= -i\omega_a(1+\beta a^\dagger a) a + [G(N)(1-i\alpha_L)-K_l(\omega)]a + \overbrace{K_c(\omega)s_\mathrm{in} + F_G}^{F_a} \\
    s_\mathrm{out}(\omega) &= K_a(\omega) a(\omega) -K_s(\omega)s_\mathrm{in}(\omega),
    \label{eq:io2}
\end{align}
where 
\begin{align}
    \begin{split}
        K_a(\omega) &= \sqrt{2\kappa_\mathrm{FSR}}\left(1-\frac{\gamma}{i\delta_d+\gamma}\right) \\
        K_s(\omega) &= 1-\frac{2\gamma}{i\delta_d+\gamma} \\
        K_l(\omega) &= \kappa_\mathrm{FSR}\left(1-\frac{\gamma}{i\delta_d+\gamma}\right)
        \\ K_c(\omega) &= K_a(\omega).
    \end{split}
\end{align}


To compute noise, we begin with the intracavity fluctuations, which are governed by the linearized system
\begin{align}
    M(\Omega)\begin{bmatrix}
    \delta a(\Omega)\\\delta a^{\dagger}(\Omega) \\ \delta N(\Omega)
    \end{bmatrix} = \begin{bmatrix}
    F_a(\Omega) \\ F_{a^\dagger}(\Omega) \\ F_N(\Omega)
    \end{bmatrix}=\begin{bmatrix}
    K_c(\omega_+)\delta s_\mathrm{in}(\Omega) + F_G(\Omega) \\ K^*_c(\omega_-)\delta s^\dagger_\mathrm{in}(\Omega) + F^\dagger_G(\Omega) \\ F_N(\Omega)
    \end{bmatrix},
\end{align}
where for a generic operator $X^\dagger(\Omega)=[X(-\Omega)]^\dagger$ follows from the definition $X^\dagger(t)=[X(t)]^\dagger$. The fluctuation matrix has columns
\begin{align}
\begin{split}
        M_{x1}(\Omega) &= \begin{bmatrix}
    -i\omega_+ + i\omega_a(1+2\beta|\alpha|^2) + K_l(\omega_+) - G(N)(1-i\alpha_L) \\ - i\omega_a\beta\alpha^{*2} \\ 2G(N)\alpha^* \end{bmatrix} \\
    M_{x2}(\Omega) &= \begin{bmatrix}
        i\omega_a\beta\alpha^2 \\ i\omega_- -i\omega_a(1+2\beta|\alpha|^2) + K_l^*(\omega_-) - G(N)(1+i\alpha_L) \\ 2G(N) \alpha \end{bmatrix}\\
        M_{x3}(\Omega) &= \begin{bmatrix}
            -G_N\alpha (1-i\alpha_L) \\ -G_N\alpha (1+i\alpha_L) \\  -i\omega+\gamma_\parallel + 2G_N|\alpha|^2
        \end{bmatrix}.
\end{split}
\end{align}
Here, $x\in [1,2,3]$ to denote the row of $M$, $\omega_{\pm}=\omega\pm\Omega$, and the steady state cavity amplitude $\alpha$ is determined through
\begin{equation}
    [i(-\omega_a(1+\beta|\alpha|^2) + \omega) + G(N)(1-i\alpha_L)- K_l(\omega)]\alpha = 0.
\end{equation}
Noise emerges from the nonzero correlators \cite{chow2012semiconductor,yamamoto1986internal}
\begin{align}
\begin{split}
    \langle F^\dagger_G(\omega) F_G(\omega')\rangle &= G(N)\delta(\omega-\omega') \\
    \langle F^\dagger_N(\omega) F_G(\omega')\rangle &= -\alpha G(N)\delta(\omega-\omega') \\
    \langle F^\dagger_N(\omega) F_N(\omega')\rangle &= [nG(N)+\gamma_\parallel N + \epsilon I]\delta(\omega-\omega')\\
        \langle s_\mathrm{in}(\omega)s^\dagger_\mathrm{in}(\omega')\rangle &= \delta(\omega-\omega'),
\end{split}
\end{align}
with $\epsilon=0$ for quiet pumping and $\epsilon=1$ for shot noise limited pumping. We can now compute the output photon noise by noting 
\begin{align}
    \begin{split}
        \delta n_\mathrm{out}(t) &= s_0^*\delta s_\mathrm{out}(t) + s_0 \delta s_\mathrm{out}^\dagger (t) \\
        \delta n_\mathrm{out}(\Omega) &= s_0^*\delta s_\mathrm{out}(\Omega) + s_0 [\delta s_\mathrm{out} (-\Omega)]^\dagger\\
        &= s_0^*K_a(\omega_+)\delta a(\Omega) + s_0K_a^*(\omega_-)\delta a^\dagger(\Omega) - [s_0^*K_s(\omega_+)\delta s_\mathrm{in}(\Omega)+s_0K_s^*(\omega_-)\delta s^\dagger_\mathrm{in}(\Omega)]
    \end{split}
\end{align}
where here $s_0(\omega)=K_a(\omega)\alpha(\omega)$ is the steady state output (propagating) amplitude and the intensity noise spectrum is given by $\langle \delta n_\mathrm{out}^\dagger (\Omega) \delta n_\mathrm{out}(\Omega) \rangle$. Spectra in the limit of nondispersive loss closely match those found by Yamamoto et al. \cite{yamamoto1986internal}.



\section{Intensity noise in QCLs with nonlinear dispersive loss}

The photon and carrier dynamics for QCLs are conventionally described using a three-level model for the carrier populations \cite{rana2002current}
\begin{align}
    \begin{split}
        \dot N_3^j &= I^j - N_3^j\left(\frac{1}{\tau_{32}}+\frac{1}{\tau_{31}}\right)-nG\left(N_3^j,N_2^j\right) +F_3^j \\
        \dot N_2^j &= \frac{N_3^j}{\tau_{32}}-\frac{N_2^j}{\tau_{21}}+nG\left(N_3^j,N_2^j\right) + F_2^j \\
        \dot n &= n\left(-\kappa(n) + \sum_{j=1}^m G\left(N_3^j,N_2^j\right)\right)+F_n,
    \end{split}
    \label{eq:qcl_eom}
\end{align}
where $N_3^j,N_2^j$ respectively denote the carrier populations in levels 3 and 2 in each gain stage $j$, $I^j$ denotes the injected current to gain stage $j$, and $\tau_{31},\tau_{32},\tau_{21}$ are the nonradiative decay time constants indicated in Fig. 4c. A linear model for the gain $G\left(N_3^j,N_2^j\right)=G_N(N_3^j-N_2^j)$ is employed. Langevin forces $F_n^j,F_3^j,F_2^j$ are added for the following noise analysis.

We can simplify the analysis by introducing $N_{2,3}=\sum_j N_{3,2}^j$ and assuming all of the gain stages are identical. Then, the dynamics for $n,N_2,N_3$ are described by a set of three coupled nonlinear equations. Note that we neglect the dynamics of $N_1$ (the carrier population in level 1) since the populations of interest $n,N_2,N_3$ form a closed system of equations. Linearizing and Fourier transforming the QCL rate equations, we find 
\[
M
\begin{bmatrix}
\delta N_3(\Omega)\\
\delta N_2(\Omega) \\
\delta n(\Omega)
\end{bmatrix} 
= 
\begin{bmatrix}
F_3\\
F_2\\
F_n - n\kappa_\omega F_\phi
\end{bmatrix}    
\]
with the fluctuation matrix
\[
M =    
\begin{bmatrix}
    -i\Omega + \gamma_{11} & -\gamma_{12} & \gamma_{13}\\
    -\gamma_{21} & -i\Omega+\gamma_{22} & -\gamma_{23} \\
    -\gamma_{31} & \gamma_{32} & -i\Omega-\gamma_{33}
\end{bmatrix},
\]
where $\gamma_{11}=nG_N+1/\tau_{32}+1/\tau_{31},\gamma_{12}=nG_N,\gamma_{13}=\gamma_{23}=G_N\Delta N,\gamma_{21}=nG_N+1/\tau_{32},\gamma_{22}=nG_N+1/\tau_{21},\gamma_{31}=\gamma_{32}=nG_N,\gamma_{33}=-n\kappa_n$ and $\Delta N=N_3-N_2$. The correlators between the Langevin forces are given by $\langle 2D_{nn}\rangle = 2G_NN_3n$, $\langle 2D_{\phi\phi}\rangle = G_NN_3/(2n)$, $\langle 2D_{22}\rangle=2G_NN_3n + N_3/\tau_{32}$, $\langle 2D_{33}\rangle = 2G_NN_3n + N_3/\tau_{32}+N_3/\tau_{31}$, $\langle 2D_{3n}\rangle = -G_N\left(N_2+N_3\right)n$, $\langle 2D_{2n}\rangle=G_N\left(N_2+N_3\right)n$, $\langle 2D_{32}\rangle = -\left(G_N(N_2+N_3)n+N_3/\tau_{32}\right)$.

In QCLs, the intensity noise is dominated by both spontaneous emission and nonradiative decay of excited carriers, whereas in typical semiconductor lasers, it is only the former that matters \cite{gensty2005intensity}. Thus, starting from the linearized matrix equations, we can approximate the DC intensity noise as
\begin{align}
    \langle \delta n^\dagger(0)\delta n(0)\rangle \approx \frac{\gamma_s^2(\gamma_{21}-\gamma_{22})^2\langle 2D_{33}\rangle + (\gamma_s\gamma_{21}-\gamma_{11}\gamma_{22})^2\langle 2D_{nn}\rangle}{\left(\gamma_s^2 \gamma_{23}+\gamma_{11}\gamma_{22}\gamma_{33}+\gamma_s(\gamma_{13}(\gamma_{21}-\gamma_{22})-\gamma_{11}\gamma_{23}=\gamma_{21}\gamma_{33})\right)^2},
\end{align}
where $\gamma_s=\gamma_{12}=\gamma_{31}=\gamma_{32}=nG_N$. In the absence of nonlinear dispersive loss, $\gamma_{33}=0$ and the DC intensity noise goes as $(\tau_s/\tau_{\mathrm{nr}})^2$ where $1/\tau_s\equiv nG_N$ is the rate of stimulated emission (per carrier) and $1/\tau_{\mathrm{nr}}$ is an effective nonradiative decay rate of the carriers. The scaling with the stimulated emission lifetime is expected given that the light approaches a coherent state as the power increases. The inverse scaling with $\tau_{\mathrm{nr}}$ reflects the fact that in QCLs, in contrast to conventional semiconductor lasers, the carrier density is not clamped above threshold. Instead, $N_2,N_3$ are dynamic and their fluctuations have fast response times, significantly affecting the intensity noise even above threshold. We also note that the fast nonradiative decay of the carriers also leads to the relaxation oscillations in QCLs being overdamped, despite increasing intensity noise. In this case, the increased intensity noise of QCLs compared to conventional lasers stems from stronger low-frequency noise arising from the unclamped carrier populations above threshold (which increase proportionately with pump current, together with the photon number). The effect of nonlinear dispersive loss is to outcompete the nonradiative decay rates to dominate the intensity noise. Thus, $|\gamma_{33}|\gg 1/\tau_{\mathrm{nr}}$ is a necessary condition for this mechanism for squeezing to be effective.

To provide analytical checks against previous theory on QCL intensity noise \cite{rana2002current}, we compute output photon noise as described above in Sec. \ref{sec:output}, agreeing qualitatively with Eq. 95 of \cite{rana2002current}.

\section{Nonlinear dispersive loss with carrier and Kerr nonlinearities}
\label{sec:carrier}

\begin{figure*}[t]
    \centering
    \includegraphics[scale=0.8]{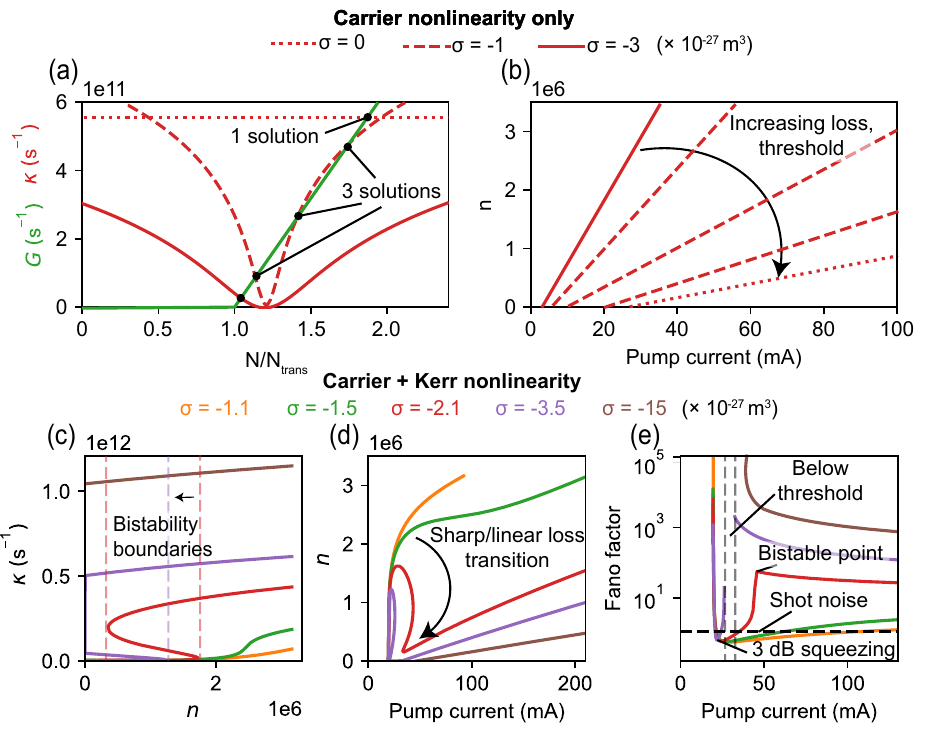}
    \caption{\textbf{Effects of carrier and Kerr nonlinearities composed with dispersive loss}. (a) In the presence of only carrier nonlinearity $\sigma$, the resonance frequency and thus loss depend ``directly'' on carrier density $N$, and steady states are set by the intersections of gain and loss. For strong $\sigma$ and low background loss $\kappa_0$, multiple steady state carrier densities $N$ can correspond to a given photon number $n$, resulting in different steady state losses (detunings from the Fano resonance). The lowest loss solution (smallest detuning) is most likely to lase, though extra solutions may be accessible by dynamic pumping schemes. (b) The schematic effect of this ``carrier bistability'' is to create multiple branches in the S-curve of different slope/threshold current. The presence of both strong carrier and Kerr nonlinearities result in the novel behaviors shown in panels (c), (d), and (e). Carrier nonlinearity causes a deformation of the intensity-nonlinear Lorentzian loss profile, eventually pinching off the ``sharp loss'' from the linear loss for sufficiently strong carrier nonlinearity (purple curve). This stems from leftward motion of the carrier bistability boundaries and creates a demarcation between linear ($F\gg 1$) and nonlinear ($F<1$) loss regimes which may be separated by a region of lasing with no stable solution. System parameters used are the same as those in Fig. 2 with $\beta=-10^{-10}$, $\kappa_0=10^{-2}\cdot\mathrm{FSR}$, and $\gamma=2\times 10^{12}$ rad/s. The magnitudes of Kerr and carrier nonlinearities taken here are comparable to what they might be in GaAs-based gain media: $\beta\sim -10^{-10}$ and $\sigma\sim -3\times 10^{-27}$ m\textsuperscript{3} (with the proviso of being taken as instantaneous and being evaluated at a single wavelength).}
    \label{fig:carrier}
\end{figure*}

In this section, we describe how the interplay of carrier nonlinearity with dispersive loss can result in unexplored ``carrier bistability'' behavior in conventional semiconductor lasers. We consider carrier nonlinear strengths $\sigma$ comparable to what they might be in, for example, GaAs-based gain media \cite{boyd2020nonlinear, said1992determination}. 

We first describe how multiple lasing steady states can exist when strong carrier nonlinearity and dispersive loss are simultaneously present. In an ordinary semiconductor laser, the ``gain equals loss'' requirement leads to a so-called ``gain clamping'' condition, wherein above threshold, the inverted carrier density is fixed at some value, regardless of the intensity (i.e. the carrier density $N$ such that $G_N(N-N_{\mathrm{trans}})=\kappa$). This is depicted in Fig. \ref{fig:carrier}a by the ``linear loss'' case which shows only a single intersection point of the carrier-dependent gain and carrier-independent loss. However, in the presence of strong carrier nonlinearity and sharply frequency-dependent outcoupling (with a Fano mirror for example), the loss of the cavity mode can depend nonlinearly on the carrier density $N$, $\kappa(\omega(N))=\kappa(N)$. As the carrier density changes, so does the cavity frequency, and hence the damping rate via the frequency-dependent mirror. The ``gain equals loss'' condition now reads $G_N(N-N_{\mathrm{trans}})=\kappa(N)$. As shown in Fig. \ref{fig:carrier}a, this leads to a situation where more than one carrier density $N$ can cause gain and loss to be equal, corresponding to multiple cavity resonance frequencies. In the case of the Fano resonance, we see that up to three different steady states are possible.

Fig. \ref{fig:carrier}b shows how this phenomenon manifests in the steady state laser behavior. The dependence of steady state intensity on the pump current is still linear, but there can be up to three independent branches, corresponding to different steady state $N$ and different lasing frequencies. For the Fano mirror example, one resonance frequency is always present such that the detuning from the Fano resonance $\Delta\approx 0$. Since this solution has lowest loss, and thus the lowest threshold, lasing will occur here by default. The other branches are also stable, but disconnected from the lowest branch. It may be possible to experimentally access these higher branches through dynamic pumping schemes which generate transients that can travel from one branch to another.


When Kerr nonlinearity is also introduced, additional phenomena appear due to the simultaneous nonlinear dependence of the damping rate on intensity and carrier number. It is important to note that the profile $\kappa(\omega)$ is unchanged, though $\kappa(n,N)$ will vary based on the nonlinear strengths. Furthermore, the gain (and thus loss) is monotonically increasing in $N$. For typical materials (and for the results presented in Fig. \ref{fig:carrier}), $\sigma<0$ increases the resonance frequency $\omega_R(n,N)$ and thus pushes lasing solutions rightward along $\kappa(\omega)$.

Consider first weak carrier nonlinearity (orange and green curves in Fig. \ref{fig:carrier}c). Then, the carrier nonlinearity can be treated as a perturbation to the initially symmetric Lorentzian loss $\kappa(n)$. On the $\kappa_\omega\equiv d\kappa/d\omega>0$ (right) branch of the dispersive loss, the carrier nonlinearity shifts the loss curve upward. On the other hand, the $\kappa_\omega<0$ (left) branch shifts downwards since an increase in $\omega$ corresponds to a point of lower loss (approaching detuning $\Delta=0$). 

Suppose we increase the carrier nonlinearity further (red curve in Fig. \ref{fig:carrier}c). For $n$ near threshold, far below the ``magic'' photon number $n_c\approx 10^6$ of lowest loss, the carrier nonlinearity pushes solutions rightward along the Lorentzian. However, the laser still lies on the $\kappa_\omega<0$ branch - the carrier nonlinearity pushes the mode closer to $n_c$, which is near zero loss and thus $N\approx N_{\mathrm{trans}}$. This yields one steady state solution. For higher $n$, near but still below $n_c$, we eventually reach an $n$ at which two solutions are possible: $N\approx N_{\mathrm{trans}}$ (lower loss) or $N > N_{\mathrm{trans}}$ (higher loss). Immediately afterwards, a third solution is possible with still higher loss/higher carrier density, phenomenologically similar to the dashed curve in Fig. \ref{fig:carrier}a. Finally, as $\kappa_\omega$ drops past the inflection point of $\kappa(\omega)$, a point corresponding to two solutions marks the end of the carrier bistability and for the largest $n$ we again obtain only one solution (the Lorentzian loss looks approximately linear).

For even stronger carrier nonlinearity (purple curve in Fig. \ref{fig:carrier}c), the carrier bistability boundaries shift leftward in photon number. Comparing the red and purple curves in Fig. \ref{fig:carrier}c, the left boundary eventually crosses zero and becomes negative, at which point the loss curve detaches into two parts: a sharp part at low loss and linear part at higher loss, separated by a range of pump currents over which no stable lasing solution occurs. When the right bistability boundary also crosses $n=0$ the sharp loss vanishes and laser operation only occurs on the linear high-loss branch (with correspondingly larger threshold currents), as shown for the brown curve in Fig. \ref{fig:carrier}c.

We now examine the effects of this carrier bistability on intensity noise. As shown in Fig. \ref{fig:carrier}e, the minimum achievable Fano factor is relatively independent of the level of carrier nonlinearity. This can be seen by noting that the first term in Eq. \ref{eq:ffactor} dominates the Fano factor at these points. However, past the sharp loss region, the linear branch created by the carrier nonlinearity possesses a higher loss that pulls the Fano factor upward for larger pump currents. For large carrier nonlinearities, the system eventually hits bistability and a region of unstable lasing, transitioning to (approximately) linear behavior again. For carrier nonlinearities much stronger than the Kerr nonlinearity (brown curve), approximately linear loss is restored as described above and no intensity noise reduction is observed. Mathematically, Eq. \ref{eq:ffactor} essentially contains a combination of dominant Kerr and dominant carrier nonlinearity terms, demarcated by pump currents smaller and larger than the Fano factor minimum/sharp loss regime, respectively.

\bibliographystyle{unsrt}
\bibliography{supp.bib}